\documentclass[10pt]{article}

\usepackage{amsmath}
\usepackage{amssymb}

\usepackage{graphicx}
\usepackage{subfigure}
\usepackage{epsfig}

\usepackage{cite}

\usepackage{color}

\usepackage[labelfont=bf,labelsep=period,justification=raggedright]{caption}

\makeatletter
\renewcommand{\@biblabel}[1]{\quad#1.}
\makeatother

\date{}

\begin{document}

\begin{flushleft}
{\Large
\textbf{Hierarchical information clustering by means of \\ topologically embedded graphs}
}
\\
\vspace{0.5cm}
Won-Min Song$^1$,
T. Di Matteo$^{1,2}$,
Tomaso Aste$^{1,3}$
\\
\vspace{0.3cm}
\bf{1} Applied Mathematics, Research School of Physics and Engineering, The Australian National University, Canberra ACT 0200, Australia.
\\
\bf{2} Department of Mathematics,  King's College London, London, WC2R 2LS, UK.
\\
\bf{3} School of Physical Sciences, University of Kent, UK.
\\
$\ast$ E-mail, Corresponding author: tomaso.aste@anu.edu.au
\end{flushleft}

\section*{Abstract}
We introduce a graph-theoretic approach to extract clusters and hierarchies in complex data-sets in an unsupervised and deterministic manner, without the use of any prior information.
This is achieved by building topologically embedded networks containing the subset of most significant links and analyzing the network structure.
For a planar embedding, this method provides both the intra-cluster hierarchy, which describes the way clusters are composed, and the inter-cluster hierarchy which describes how clusters gather together.
We discuss performance, robustness and reliability of this method by  first  investigating several artificial data-sets, finding that it can outperform significantly other established approaches.
Then we show that our method can successfully differentiate meaningful  clusters and hierarchies in a variety of real data-sets.
In particular, we find that the application to gene expression patterns of lymphoma samples uncovers biologically significant groups of genes which play key-roles in diagnosis, prognosis and treatment of some of the most relevant human lymphoid malignancies.



%
\section*{Introduction}

Filtering information out of complex datasets is becoming a central issue and a crucial bottleneck in any scientific endeavor.
Indeed, the continuous increase in the capability of automatic data acquisition and storage is providing an unprecedented potential for science.
However, the ready accessibility of these technologies is posing new challenges concerning the necessity to reduce data-dimensionality by filtering out the most relevant and meaningful information with the aid of automated systems.
%
%
In complex datasets information is often hidden by a large degree of redundancy and grouping the data into clusters of elements with similar features is essential in order to reduce complexity \cite{Jain1999}.
However, many clustering methods require some  \emph{a priori} information and must be performed under expert supervision.
The requirement of any prior information is a potential problem because often the filtering is one of the preliminary processing on the data and therefore it is performed at a stage where very little information about the system is available.
Another difficulty may arise from the fact that, in some cases, the reduction of the system into a set of separated local communities may hide properties associated with the global organization.
For instance, in complex systems,  relevant features are typically both local and global and different levels of organization emerge at different scales in a way that is intrinsically not reducible.
We are therefore facing the problem of catching simultaneously two complementary aspects: on one side there is the need to reduce the complexity and the dimensionality of the data by identifying clusters which are associated with local features; but, on the other side, there is a need of keeping the information about the emerging global organization that is responsible for cross-scale activity.
It is therefore essential to detect clusters together with the different hierarchical gatherings above and below the cluster levels.
In the literature there exist several methods which can be used to extract clusters and hierarchies \cite{Jain1999,McQueen1967,Xu2005} and the application to biology and gene expression data has attracted a great attention in recent years \cite{Eisen98,Rocke09,Quackenbush2001,Rivera10}.
However, in these established approaches, to extract discrete clusters, one must input some a priori information about their number or define a thresholding value.
This introduces other potential difficulties because complex phenomena are often associated with multi-scaling signals which cannot be trivially thresholded.
In this paper, we propose an alternative method that overcomes these limitations providing both clustering subdivision and hierarchical organization without the need of any prior information, without demanding supervision and without requiring thresholding.

In recent years, several network based approaches have been proposed to describe complex data-sets and applied to several fields from biology \cite{Jonsson2006,Goh07} to social  and financial systems \cite{Girvan2002,Kitsak2010}.
Indeed, networks naturally reflect in their set of vertices the variety of elements in the system, they reflect in their edges the plurality of the interrelations between elements and they encode in their dynamics the complex evolution and adaptation of the system \cite{Amaral00,Garlaschelli07,Calrarelli07,Buldyrev10,Hooyberghs10}.
In this paper we apply the network paradigm to the study of complex data-structures.
In our approach a graph with constrained complexity is built by means of a deterministic construction inserting recursively the most relevant links.
In this construction, complexity is constrained by embedding the graph on an  hyperbolic surface of genus $g$ (where the genus is the number of handles of the surface) \cite{Aste2005,Tumminello2005}.
The Ringel-Youngs theorem ensures that for $n$ vertices the complete graph, $K_n$, can be always embedded on a surface with large enough genus ($g \simeq O(n^2)$) \cite{Ringel74}.
Any graph is a sub-graph of $K_n$ and therefore any graph can be embedded on a surface.
In this paper we are interested in the limit where graphs are sparse and they are embedded on simple surfaces.
The simplest case is $g=0$ and the resulting graph is called Planar Maximally Filtered Graph (PMFG) and it is a triangulation of a topological sphere.
Topologically embedded graphs on planar surfaces ($g=0$) have a relatively small number of edges ($O(n)$) but they have high-clustering coefficients, they can display various kinds of degree distributions, from exponential to power-law tailed, and they can be used as a platform for modeling other systems  \cite{Aste2005,Andrade2005,DiMatteoWealth04,DiMatteoInFlow05,Pellegrini2007}.
It has been shown that PMFG graphs are efficient filtering tools having topological properties associated to the properties of the underlying system \cite{Tumminello2005,DiMatteo10}.
This makes the PMFG a desirable tool to extract clusters and hierarchies from complex data-sets.

The general idea at the basis of our method is to use the topological structure of PMFG graphs to investigate the properties of the data-sets.
A detailed description of our clustering and linkage procedure is reported in the Methods section.
For brevity, in the rest of the paper, we will refer to our clustering and linkage method as the \emph{DBHT technique}.

\section*{Results}
In this section, we apply the DBHT technique to various data sets ranging from artificial data with known clustering and hierarchical structures to real gene expression data.
Comparisons are made between the results retrieved by the DBHT technique and some of state-of-the-art cluster analysis techniques such as k-means++\cite{Arthur2007}, Spectral clustering via Normalized cut on k-nearest neighbor graph (kNN-Spectral) \cite{Shi2000,Luxburg2007}, Self Organizing Map (SOM) \cite{Kohonen2001} and Q-cut \cite{Ruan2010}.
Let us here stress that all these techniques --except DBHT-- are non-deterministic and require some \emph{a priori}  information in order to setup the initial parameters.
To compare with the  DBHT technique, we run the other techniques for a broad range of parameters and pick the set of parameters that are best performing in average.
This is an important negative bias against the  DBHT technique that however, as we shall see shortly, still outperforms consistently the  state-of-the-art counterparts.
We also tested the capability of DBHT technique to correctly detect the hierarchical organization by applying it to known synthetic datasets and comparing the results with the outcomes from average and complete linkage techniques.
Furthermore, we explored  the meaningfulness of the hierarchical gathering of clusters and the significance of their subdivision in sub-clusters by looking at the functional properties of these gatherings and splittings in real datasets.

\subsection*{Tests DBHT clustering on synthetic data}
We have evaluated performance of the clustering techniques by comparing their outcomes with the known artificial clustering structure by using a popular external validity index: the adjusted Rand index  \cite{Hubert1985} which returns 1 for a perfect match and in average 0 for a random guess.
Specifically, we have generated correlated data-series by using a multivariate Gaussian generator (MVG) \cite{mvg} that produces $N$ stochastic time series $y_i(t)$ of length $T=10\times N$  with zero mean and Pearson's cross-correlation matrix $R$ that approximates an input correlation structure $R^*$ which is a bolck-diagonal matrix where the blocks represent the clusters and may have different sizes.
The matrix  $R^*$ has all ones on the diagonal, it has zero correlations outside the blocks  ($\rho^{ou*}=0$) and it has a correlation value $\rho^{in*}$ inside the blocks.
Furthermore, we added a number $N_{ran}$ of random correlations unrelated to the cluster structure.
We have also generated multivariate Log-Normal distributions  by taking the exponential of MVG series generated by using reference correlation $R^*_{log}$ which is devised to retrieve the correct approximation of  $R^*$  with log-normal statistics \cite{logMVG}.
To these correlated series we have added a noise  $\eta_i(t)$  obtaining $y_i'(t)=y_i(t)+c \sigma_i \eta_i(t)$, where $\sigma_i$ is the standard deviation of $y_i(t)$ and $c$ is a constant that can be used to tune the relative amplitude of noise.
We have tested  normally distributed ($p(\eta) \propto \exp(-\eta^2/2)$), log-normally distributed ($p(\eta) \propto \exp(-\log(\eta)^2/2)$) or power-law distributed ($p(\eta) \propto 1/\eta^{\alpha+1}$) noises.
We have used different values for the relative amplitude of noise $c$ and, in the case of power-law distributed noise, we have also varied the exponent $\alpha$.
By increasing the effect of noise and/or the number of random elements, the Pearson's cross-correlation matrix $R$ passes from a very well defined structure similar to $R^*$ to a less defined structure where the difference between the average measured intra- and inter-cluster correlations in $R$,  $\left< \rho^{in} \right>-\left< \rho^{ou} \right>$, becomes negligible.

\begin{figure}[ht]
\centering
\includegraphics[scale=0.4,angle=0]{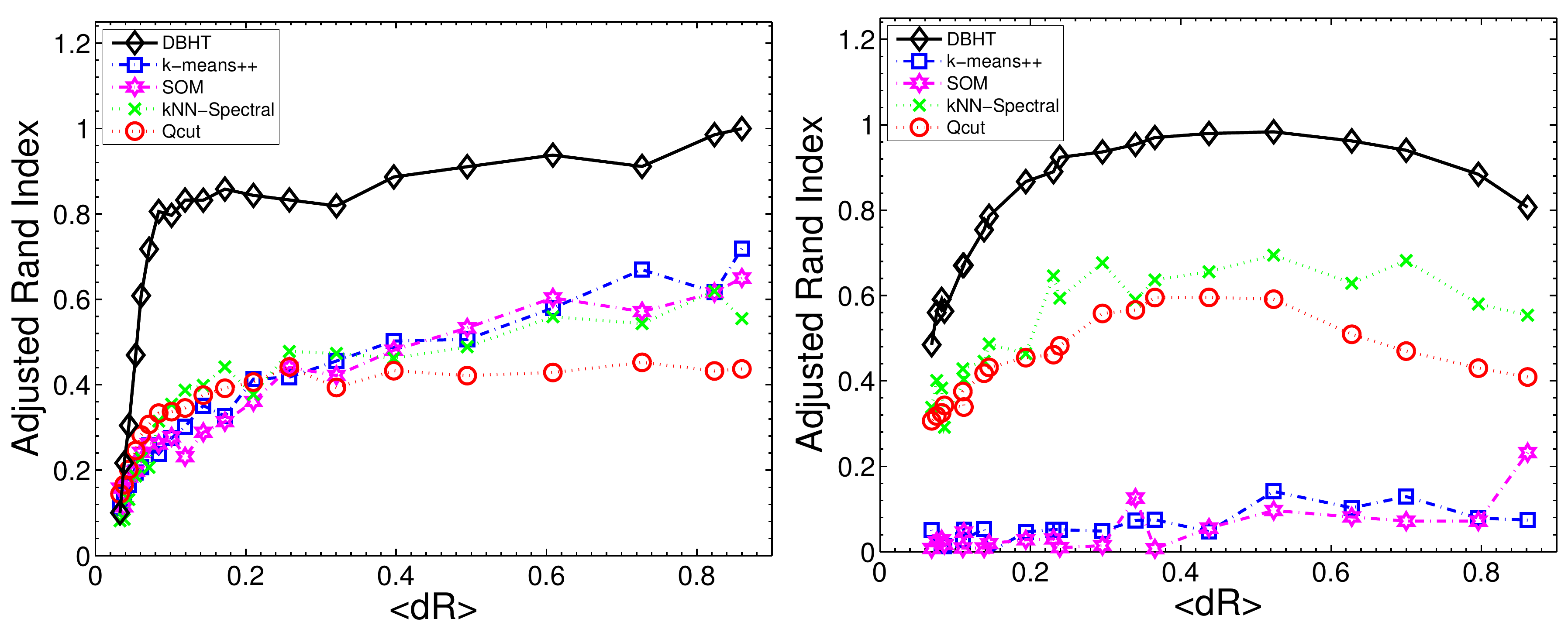}
\caption{\label{CaseSpecificPaper}
Demonstration that the DBHT technique can outperform other state-of-the-art clustering techniques, namely: k-means++\cite{Arthur2007}, Spectral clustering via Normalized cut on k-nearest neighbor graph (kNN-Spectral) \cite{Shi2000,Luxburg2007}, Self Organizing Map (SOM) \cite{Kohonen2001}, and Q-cut \cite{Ruan2010}.
This figures reports the adjusted Rand index \cite{Hubert1985} for the comparison between the the `true' partition embedded in the artificially generated data and the partition retrieved by the clustering methods.
In these examples we have eight clusters of size 5 elements and one cluster of size 64 elements with $\rho^{in*}=0.9$, $\rho^{ou*}=0$ and $N_{ran}=25$.
The plots report average values over a set of the 30 trials.
The horizontal-axis reports the gap between average intra- and inter-cluster correlations $dR = \left< \rho^{in} \right>-\left< \rho^{ou} \right>$ that becomes smaller when the noise $c$ increases.
{\bf (a)} Normally distributed correlated datasets with added Normal noise with $c$ varying from 0 to 4. 
{\bf (b)} Log-Normally distributed correlated datasets with added power law noise with $\alpha=1.5$ and $c$ varying from 0 to 0.1. 
}
\end{figure}

Figure \ref{CaseSpecificPaper} compares the performance of the DBHT technique with k-means++, SOM, kNN-Spectral and Q-cut for correlated synthetic datasets consisting of 129 data series generated both with normal and log-normal statistics, with normal or power law noise with $\rho^{in*}=0.9$, $\rho^{ou*}=0$ and $N_{ran}=25$.
This example refers to a rather extreme case where the clusters have highly dis-homogeneous sizes with one large cluster with 64 elements and eight clusters with 5 elements each.
As one can see from Fig. \ref{CaseSpecificPaper} in this case the  DBHT technique is strongly outperforming the other methods.
In the supporting information, we report on a large number of cases where we demonstrate that consistently the DBHT technique is better, or at least equivalent, to the best performing counterparts for a very broad range of combinations of different kinds of artificial data.
Let us here note that stochastic techniques such as k-means++ and SOM are particularly sensitive to noise distributions and tend to perform poorly with fat-tailed distributed noise.
On the other hand,  the Qcut technique carries an inherent resolution limit that over-shadows small clusters \cite{Fortunato2007}.
The DBHT technique instead is less affected by these factors and it consistently delivers good performances for across the range of parameters.

\begin{figure}[!ht]
\centering
\includegraphics[width=0.9\columnwidth]{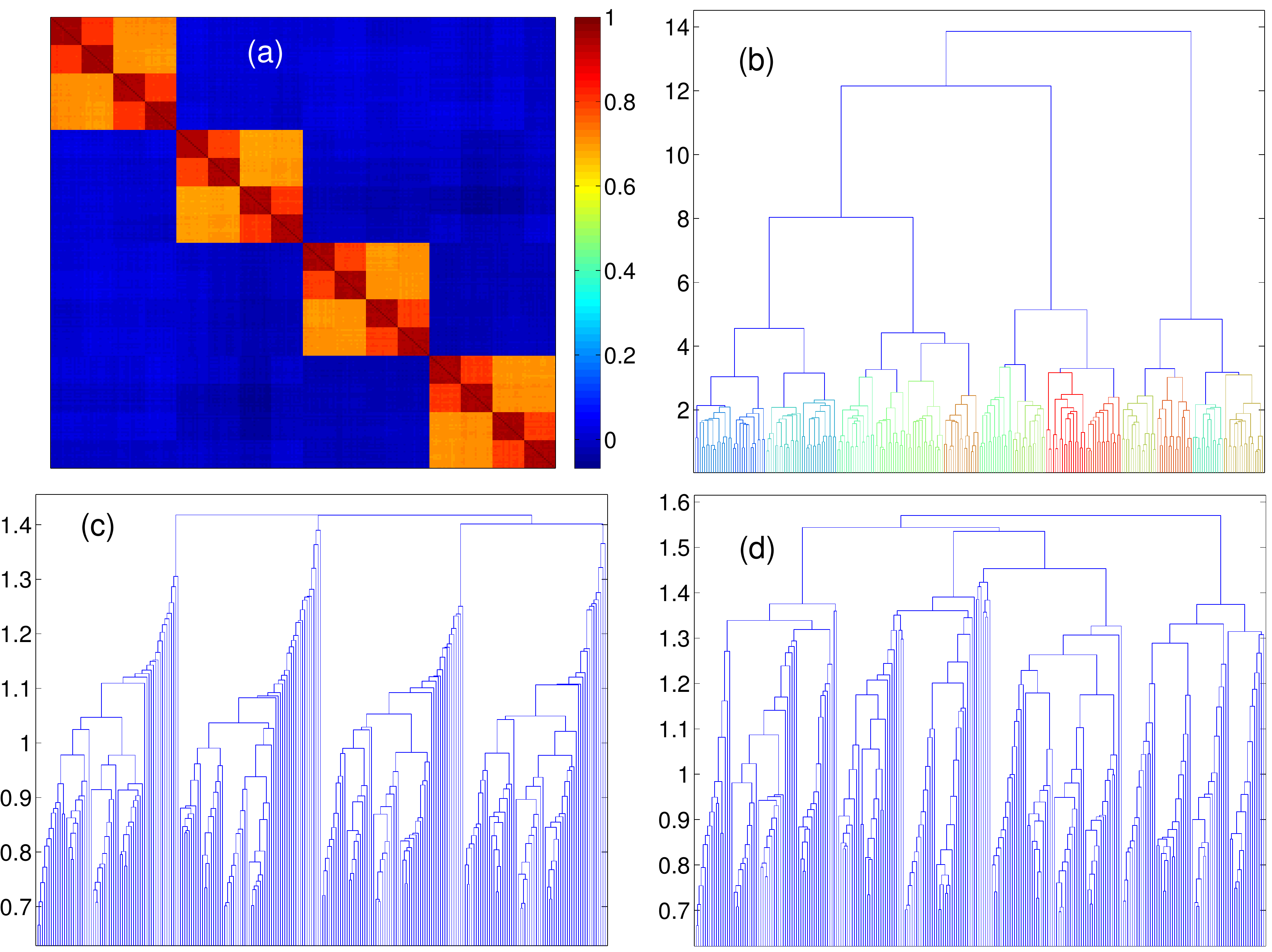}
\caption{ \label{f.4x}
Demonstration that the DBHT technique can detect clusters at different hierarchical levels outperforming other established linkage methods.
The synthetic data are generated via multivariate Gaussian with added power law noise with exponent $\alpha = 1.5$ and $c=0.1$.
{\bf (a)} Input correlation $R^*$ for a synthetic data structure with nested hierarchical clustering with 4 `large' clusters, containing 8 `medium' clusters, containing 16 `small' clusters.
{\bf (b)} Dendrogram associated with the DBHT hierarchical structure.
{\bf (c)} Dendrogram associated with the Average linkage.
{\bf (d)} Dendrogram associated with the Complete linkage.
}
\end{figure}

\subsection*{Tests DBHT hierarchy on synthetic data}
We have tested the capability of the DBHT technique to detect hierarchies by simulating data with hierarchical structure such that smaller clusters are embedded inside larger clusters making a nested structure with different intra-cluster correlation.
An example is shown in Fig.\ref{f.4x}(a) where we report an input correlation $R^*$ which is a nested block-diagonal matrix with zero inter-cluster correlation and with a structure of 4 `large' clusters (64 elements each) with intra-cluster correlation of $\rho^{in*}_1=0.7$.
Each of the large clusters contains inside two `medium' clusters (8 in total with 32 elements each) with $\rho^{in*}_2=0.8$ that contain inside two `small' clusters (16 in total with 16 elements each) with $\rho^{in*}_3=0.95$.
We have simulated 30 different sets of  data series of length $T=10\times N$ by using MVG from  $R^*$  with added power law noise with $\alpha = 1.5$ and $c=0.1$.
We have tested the efficiency of the DBHT technique by moving through the hierarchical levels varying the number of clusters from only one at the top hierarchy to the number of elements at the lowest hierarchy.
Fig.\ref{f.4x}(b) shows the dendrogram retrieved with the DBHT technique.
By following the hierarchy from top to bottom, one can see that a structure with 4 main clusters rapidly emerges and its partition coincides exactly with the `true' partition in $R^*$.
Then these clusters correctly split into two parts each making 8 clusters in total scoring a value of 0.97 for the adjusted Rand index with respect to the `true' partition at this level. 
Finally, these 8 clusters  split again producing a partition that has an adjusted Rand index of 0.94 with respect to the `true' partition at this level.
The partition into discrete clusters identified by the DBHT is almost identical with this last one having 17 clusters instead of the 16 `true' clusters and achieving also an adjusted Rand index of 0.94 (see supporting information).
One can see from Fig.\ref{f.4x}(c,d) that, instead, the complete and average linkages give a less clear hierarchical structure.
Several other examples are reported in the supporting information.
The better performances of the DBHT technique over linkage methods can be explained by the fact that linkage techniques suffer from the greedy nature of the algorithm, where a misclassification of an element in an early stage of clustering can never be remedied \cite{Jain1999,Xu2005}.
The rate of misclassification depends on the type of linkage distance, with the average linkage optimized for isotropic clusters, and complete linkage optimized for compact and well-defined clusters.
On the other hand, DBHT hierarchy is based on a combination of linkage distance and topological constraints at multiple hierarchical levels: bubbles, clusters, bubble tree.
This reduces the error rate with respect to the complete linkage distance.

\begin{figure}[!ht]
\centering
\includegraphics[width=0.95\columnwidth]{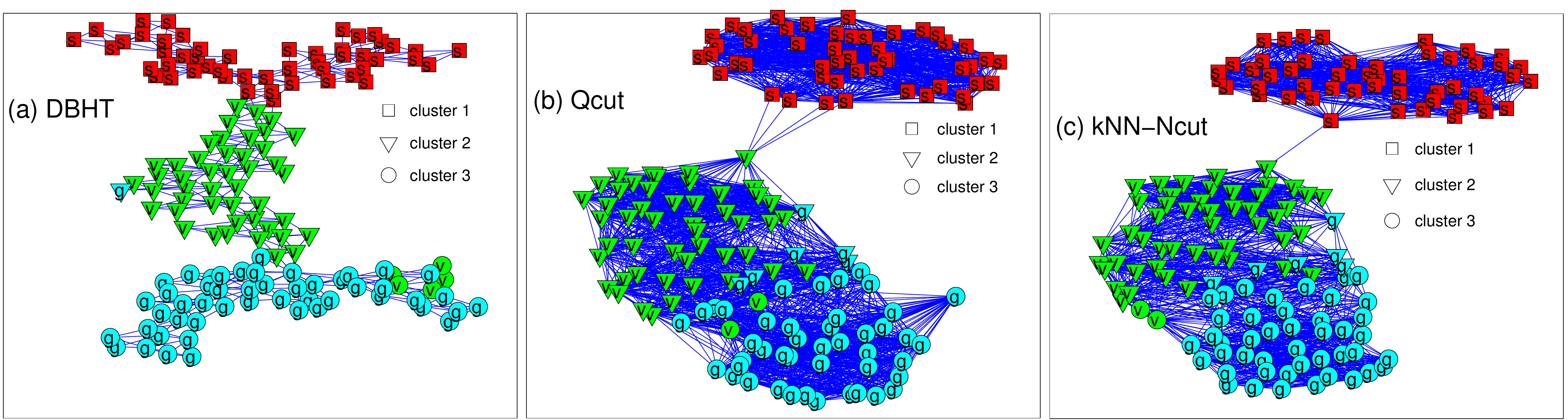}
\caption{\label{IrisGraphs}
Comparison beween the clustering obtained via (a) DBHT technique, (b) best Qcut and (c) best kNN-Spectral on iris flower data set from Fisher \cite{Fisher36}.
The labels inside the symbols correspond to the three different types of flowers: (s) Iris Setosa; (v) Iris Versicolour; (g) Iris Virginica.
The shapes of the symbols correspond to the clusters retrieved by the different clustering techniques.
}
\end{figure}

\subsection*{Application of DBHT technique to Fisher's Iris Data  }
One of the typical benchmark referred in clustering analysis literature is the iris flower data set from Fisher \cite{Fisher36}.
Briefly, the data set contains the measure of four features
(i) sepal length;
(ii) sepal width;
(iii) petal length;
(iv) petal width,
for 50 iris plants from three different types of iris, namely
(1) Iris Setosa;
(2) Iris Versicolour;
(3) Iris Virginica.
The data set is available from UCI Machine Learning Repository website \cite{IrisWebsite}.
It is known that, the clustering structure of the data set linearly separates one type of Iris from the other two.
The remaining two types are instead not linearly separable and their subdivision is a classical challenge for any clustering technique \cite{IrisWebsite}.
Here, in order to compute clustering and hierarchies  we have used the pair-wise Euclidean distance $\mathbf{D_{euc}}(i,j)=\|x_i-x_j\|$ as dissimilarity matrix and $\mathbf{R_{euc}}(i,j)=\exp{(-\frac{\|x_i-x_j\|^2}{2\sigma^2})}$  as similarity matrix \cite{Luxburg2007}, where $\sigma$ is the standard deviation of $\mathbf{D_{euc}}(i,j)$ for all pairs of $(i,j)$.
From these measures, we directly computed clustering and hierarchies via DBHT technique obtaining the graph structure shown in Fig.\ref{IrisGraphs}(a) where one can see that all the three iris types are rather well separated occupying different parts of the graph.
By extracting three clusters form the DBHT hierarchy we observe that the first flower type  (Iris Setosa) is fully separated and the other two are rather well divided with only a few misplacements.
The DBHT results are compared with other two graph-based techniques, Qcut and kNN-Spectral techniques computed using $\mathbf{R_{euc}}$ for a range of $kNN=2,\ldots,(N-1)$.
These methods ar non deterministic and we retained only the best partitions which give the highest adjusted Rand score which are shown in Fig.\ref{IrisGraphs}(b,c).
We can observe that Qcut and kNN-Spectral  techniques provide a poorer separation of the last two flower tipes (Iris Versicolour and Iris Virginica).
This is quantified by the adjusted Rand index computed by comparing with the true partition that gives 0.89 for DBHT and 0.85 for both Qcut and kNN-Spectral.
Indeed, these last two techniques both misplace 8 elements of the two groups whereas DBHT misplaces only six.
Other two clustering techniques, k-means++ and SOM, have been run over 30 iterations with a input number of clusters $k=3$, yielding to poorer partitions with the largest adjusted Rand indexes respectively of 0.73 and 0.80 which are well below the score achieved by the DBHT technique.

\subsection*{Application of DBHT technique to gene expression data set from human cancer samples}
We have applied the DBHT technique to analyze gene expression data sets collected by Alizadeh \emph{et al}\cite{Alizadeh00} concerning 96 malignant and normal lymphocyte samples belonging to the three most relevant adult lymphoid malignancies, namely: Diffuse Large B-Cell Lymphoma (DLBCL); Follicular Lymphoma (FL); Chronic Lymphocytic leukemia (CLL); together with other 13 kinds of samples from normal human tonsil, lymph node, Transformed Cell Line, Germinal Centre B, Activated Blood B, and Resting Blood B.
This data set has already served as a benchmark to evaluate performance of clustering techniques on gene expression data \cite{Wang2002,Ruan2010} and this is why we have chosen to test our method on this referential dataset.
Patients with DLBCL  cancer type have variable clinical courses and different survival rates and there are strong indications that  DLBCL classification includes more than one disease entity \cite{Alizadeh00}.
The challenge for a clustering algorithm is therefore to analyze the DLBCL genetic profiles and individuate different subtypes of DLBCL to be associated with different clinical courses.
Indeed, various studies have attempted to highlight genetically significant genes that can be of clinical significance to improve the DLBCL patients' diagnosis and clinical treatment \cite{Alizadeh00,Abramson2005,Lenz2008,Wada2009,Johnson2009,Zhao2007,Filipits2002}.
In particular, it is understood that DLBCL is a very heterogeneous type of Lymphoma and there are at least three distinct subtypes which differ in treatment methods for improved survival of the patients \cite{Alizadeh00,Abramson2005,Chen2008}.

\begin{figure}[ht!]
\centering
\includegraphics[width=0.8\columnwidth]{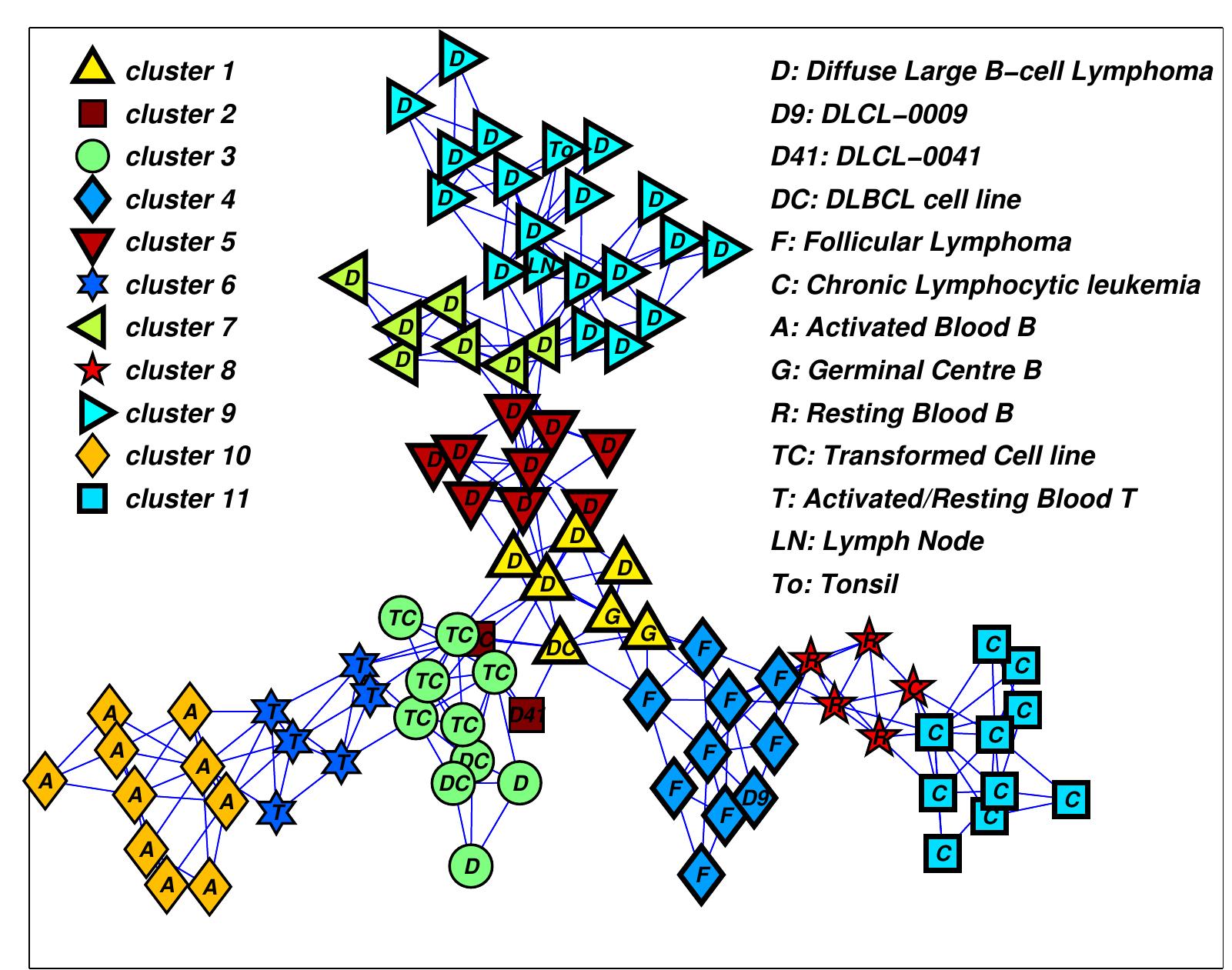}
\caption{
Sample-cluster structure for 96 malignant and normal lymphocyte samples from Alizadeh \emph{et al} 2000 \cite{Alizadeh00}, the labels inside the symbols correspond to the different sample types as listed in the legend.
The DBHT technique retrieves 11 sample-clusters here represented with different symbols (see legend).
The underlying network is the PMFG from which the clustering has been computed.
\label{DLBCL_PMFG}}
\end{figure}

\begin{table}[ht]							
\centering									
\begin{tabular}{|c|c|c|c|c|}									
\hline									
	&	Sample Cluster `1'	&	Sample Cluster  `5'	&	Sample Cluster `7'	&	Sample Cluster `9'	\\
\hline									
Cluster Size	&	7	&	9	&	7	&	20	\\ \hline	
\# of DLBCL	&	4	&	9	&	7	&	17	\\ \hline	
\# Survived over 5yr    	&	3 (100\%)	&	5 (56\%)	&	1 (14\%)	&	5 (29\%)	\\ \hline	
\# Died in 5yr	&	0	&	4	&	6	&	12	\\ \hline									
\end{tabular}									
\caption{\label{SurvTable}
Survival rates of cancer patients with DLBCL  type of Lymphoma.
The patients are divided in four groups corresponding to the four sample-clusters containing the DLBCL samples (see Fig.~\ref{DLBCL_PMFG}).
}
\end{table}

We have first applied the DBHT technique on the gene expression data by using Pearson's correlation as similarity measure, and correlation distance as the dissimilarity measure.
The DBHT clustering yielded to 11 sample-clusters, which are shown in Fig.~\ref{DLBCL_PMFG}.
One can immediately note that all FL  samples are gathered together in one cluster that also contains the DLCL-0009 sample which it has also been associated to FL in other studies on the same data \cite{Alizadeh00,Ruan2010}.
Transformation of FL to DLBCL is common \cite{Lossos2002}, and this cluster suggests that DLCL-0009 may have derived from FL, sharing therefore common gene expression patterns.
We also observe in Fig.~\ref{DLBCL_PMFG} that all, except one, the CLL samples occupy a single cluster.
The missing CLL sample is attached to this cluster and it is included in a cluster containing Resting Blood B  samples which have indeed similar expressions patterns and clinical similarity to CLL and are often merged together by other clustering techniques \cite{Ruan2010}.
DLBCL cancer types appear in four different sample-clusters which are however lying together in a branch of the PMFG graph.
Significantly, these clusters also include some other GCB-like samples.
Remarkably, if we look at the patient survival rates (Table~\ref{SurvTable}), we see that these four sample-clusters are extracting DLBCL cancer subtypes with very different clinical courses.
Indeed, if we consider separately the patients with DLBCL type of Lymphoma accordingly with the subdivision into the four sample-clusters `1', `5', `7' and `9' (from bottom to top of the Fig.~\ref{DLBCL_PMFG}), they respectively have survival rates 100\%, 56\%, 15\% and 29\% (see Table \ref{SurvTable} for details).
In the work of Alizadeh \emph{et al}\cite{Alizadeh00} survival rate differentiation in DLBCL patients was associated with two main cancer subtypes, namely GCB-like and ABC-like, with the latter considered more fatal that the former.
We can note that, in our clustering, sample-cluster `1' contains GCB-like DLBCL, and it also includes other GCB samples such as tonsil GCB, tonsil GC fibroblast, and high survival rates are common in GCB-like cancer types (see Fig. 10 in supporting information).
Cluster `5' is also characterized by GCB-like DLBCL samples, however its proximity to ABC-like clusters (see Fig. 10 in supporting information), may be the clue to relatively low survival rate in comparison to cluster `1'.
Cluster `9' is characterized by a majority of ABC-like DLBCL to which we may attribute its relatively low survival rate \cite{Alizadeh00}. On the other hand, cluster `7', which shows a surprisingly low survival rate, has instead a significant number of GCB-like DLBCL samples, this might signal the existence of another relevant DLBCL subtype.

\begin{table}[ht]
\centering
\begin{tabular}{|c|c|c|c|c|c|c|c|}
\hline 
 & GCB & LyN & PBC & Pr & TC & ABC\\
\hline
Sample Cluster `1'&\textbf{61/0}&0/2&27/0&\textbf{115/0}&1/15&4/12\\
\hline
Sample Cluster `2'&2/0&0/2&0/2&7/3&0/1&0/3\\
\hline
Sample Cluster `3'&0/35&2/37&0/15&259/0&0/38&4/3\\
\hline
Sample Cluster `4'&83/0&0/97&48/0&1/193&3/12&0/37\\
\hline
Sample Cluster `5'&\textbf{21/2}&97/0&7/3&\textbf{119/0}&2/4&0/11\\
\hline
Sample Cluster `6'&7/27&1/47&0/61&6/126&86/0&32/0\\
\hline
Sample Cluster `7'&4/6&\textbf{111/0}&0/24&1\textbf{7/4}&14/3&13/1\\
\hline
Sample Cluster `8'&0/2&0/41&17/1&0/199&6/4&2/7\\
\hline
Sample Cluster `9'&1/13&133/0&7/1&\textbf{70/0}&14/4&24/2\\
\hline
Sample Cluster `10'&0/37&3/48&1/14&44/68&1/20&61/0\\
\hline
Sample Cluster `11'&20/43&0/110&27/12&0/303&20/16&1/56\\
\hline 
\end{tabular}
\caption{\label{RegulationTable}
Number of up-regulated (on the left) and / down-regulated (on the right) expression profiles for each group of clones with known physiological roles as reported in Ref.~\cite{Alizadeh00}.
The sample-cluster labels are as in Fig.~\ref{DLBCL_PMFG}.
Some significant up-/down-regulation patterns, commented in the text, are highlighted by boldface font.
}
\end{table}

In order to functionally validate these sample-clusters, we have analyzed the expression profiles for 6 groups of genetic clones with known physiological roles, namely:
GCB- Germinal Center B cell (111 clones), LyN- Lymph Node (136 clones), PBC- Pan B Cell (81 clones), Pr- Proliferation (312 clones), TC- T Cell (111 clones) and ABC- Activated B Cell (86 clones) \cite{Alizadeh00}.
The significance of regulation patterns has been evaluated by one-tailed T tests with cut-off p-value of 0.01.
The number of up-/down-regulated profiles for each group of clones is shown in Table~\ref{RegulationTable}.
Significant up-/down-regulation patterns of the expression profiles in the sample-clusters reflect the biological relevance the group of gene-clones in each sample-cluster.
We first observe that sample-clusters containing DLBCL cancer types (e.g. cluster `1', `5', `7' and `9') distinguish from other samples by up-regulating more clones from Pr, hence reflecting higher proliferative index.
Sample clusters associated to DLBCL are also differentiating among themselves, for instance, sample-clusters `1' and `5' both up-regulate GCB clones but they differ significantly in the up-regulation of LyN clones, supporting the subdivision of GCB-like DLBCL by these sample clusters. Similarly, sample-cluster `7' shows a unique expression signature that highlights a strong up-regulation of LyN clones in comparison to other clones. Given that this sample-cluster is a mixture of ABC-like and GCB-like DLBCLs,  and it shows distinctively low survival rate, this again suggests that sample-cluster `7'  is a different subtype of DLBCL outside of GCB-/ABC-like classification. Overall, these results indicate that DBHT clustering technique is able to reveal a meaningful classification of biologically significant DLBCL subtypes which is richer than what proposed in the original study by Alizadeh \emph{et al} \cite{Alizadeh00}.

\begin{figure}[ht!]
\begin{center}
\includegraphics[width=0.6\columnwidth]{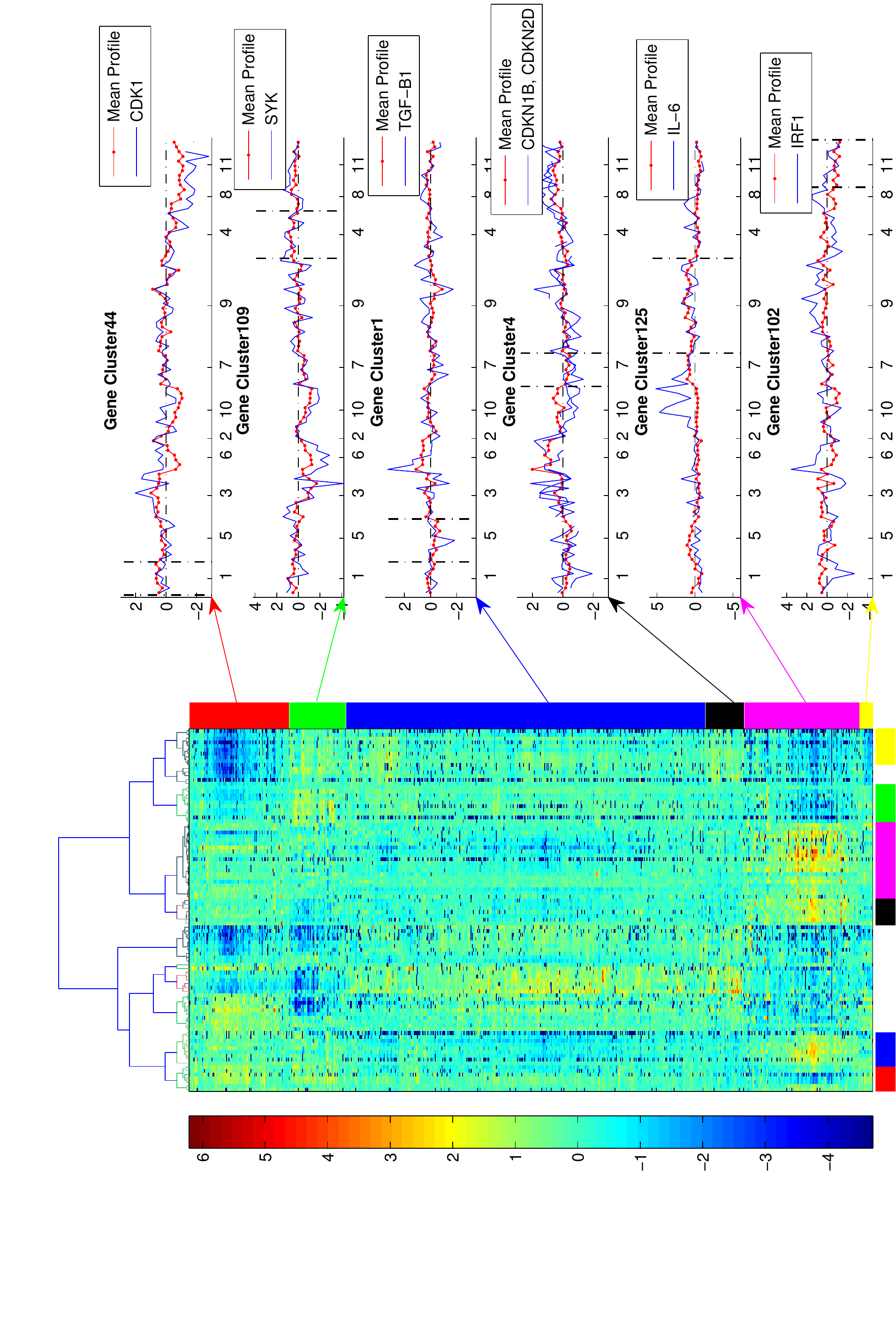}
\end{center}
\caption{\label{GeneClusters_AllSampleClusters}
\emph{Left}: Heat map of gene expression profiles for the six significant clusters of genes.
Each row represents the expression profile from a clone, and each column represents a sample.
The samples are organized according to the DBHT hierarchy as shown on the dendrogram on the top.
Significant gene-clusters are highlighted with different colors as follows (from top to bottom, colours online):
Red - gene-cluster `44' (significant for sample-cluster `1');
Green - gene-cluster `109' (significant  for sample-cluster `4');
Blue - gene-cluster `1' (significant  for sample-cluster `5');
Black - gene-cluster `4' (significant  for sample-cluster `7');
Magenta - gene-cluster `125' (significant  sample-cluster `9');
Yellow - gene-cluster `102' (significant for sample-cluster `11').
The same color scheme is used on the bottom of the heatmap to denote the corresponding sample-clusters.
\emph{Right}: Mean expression profile for each gene-cluster together with the expression profiles of note-worthy gene for each sample-cluster.
The x-axes report the gene clusters and the boundaries of the relevant sample-cluster for each gene-cluster are indicated with the vertical dashed lines.
}
\end{figure}

Let us now move a step further and use the DBHT technique to identify significant groups of genes that are of relevance for particular cancer samples.
Indeed, an accurate identification of significant genes is crucial in treating the tumor cells as there are a large number of different genetic mechanisms from which these tumor cells originate, hence they require different treatments \cite{Lam2008,Coffey2009}.
We have therefore performed a two-way clustering: on the samples and genes simultaneously.
In this way, we can cross-tabulate the samples against genes obtaining a simple and effective picture of significant gene expression patterns.
Let us note that with conventional clustering techniques, the two-way clustering adds another dimension of complexity.
Indeed, samples and gene expression profiles have different dimensions and scales and therefore it is necessary to tune the clustering parameters separately for each clustering way.
On the other hand, the DBHT technique has no adjustable parameters and it is deterministic providing therefore a unique cross classification without any increase in complexity.
The DBHT technique identifies  180  gene-clusters from which we have extracted 6 clusters which are significantly differentiating for sample-clusters associated to FL, CLL and DLBCL, accordingly with a p-value threshold of 0.01 with Bonferroni correction.
The expression profiles of these significant gene-clusters are reported in Fig.~\ref{GeneClusters_AllSampleClusters}.
We have then validated functional significance of these gene-clusters by performing a gene-ontology (GO) analysis to identify significant GO terms for biological processes \cite{Bingo}.
(See supporting information for the statistical analysis methods and GO results.)
Let us here report on some relevant genes, from each of the 6 significant gene-clusters, selected by choosing the most frequently appearing genes in the GO terms.
Interestingly, these genes reveal some of biologically significant mechanisms that regulate growth of tumor cells, and that affect survival of respective lymphoma malignancy.
In particular:
\begin{itemize}
\item Gene cluster `44' (significant for sample-cluster `1'):
This gene-cluster is up-regulated for sample-cluster `1' in comparison to the expressions in other sample-clusters associated to lymphoma. Significantly, one of its key genes is CDK1, which is a key player in cell cycle.
It has been indicated that over-expression of CDK1 is common in DLBCL cancer types, and it is therefore a potential therapeutic target \cite{Zhao2009}.
\item Gene cluster `4' (significant for sample-cluster `4'):
This gene-cluster particularly expresses for sample-cluster `4', which consists mostly of FL samples.
Among the genes in this gene-cluster there is SYK which -indeed- has been indicated as a promising target gene for antitumor therapy for treating FL, where inhibition of SYK expression increases the chance of survival \cite{Leseux2006}.
\item Gene cluster `1' (significant for sample-cluster `5'):
Gene cluster 1 is particularly down-regulated for sample-cluster `5'.
This gene-cluster contains TGF-B1 which is a well-known transcription factor to regulate proliferation, in particular a negative regulator of B-cell lymphoma which induces apoptosis of the tumor cells via NF-$\kappa$B/Rel activity \cite{Arsura1996}.
This suggests that suppression of the tumor cells by TGF-B1 would be lessened in sample-cluster `5' due to the down-regulation, and this may contribute to the decreased chance of survival observed in sample-cluster `5' in comparison to that of sample-cluster `1'.
\item Gene cluster `4' (significant for sample-cluster `7'):
This gene-cluster is slightly down-regulated for sample-cluster `7', and GO analysis extracts two genes, CDKN1B/p27$^{Kip1}$ and CDKN2D/p19, which are key tumor suppressor genes for aggresive neoplasms \cite{Kamijo1997,Seki2003}.
The inhibited tumor suppressive role of these genes might have led to aggressive growth of tumor cells suggesting a plausible explanation for the poorest survival rate, observed for sample-cluster `7', with respect to the other DLBCL sample-clusters (see Table.\ref{SurvTable}).
Indeed, it has been suggested that p27 is associated to lymphomagenesis through Skp2 \cite{Seki2003} and Skp2 has been indicated as an independent marker to predict survival outcome in DLBCL \cite{Seki2003,Saez2004}.
\item Gene cluster `125' (significant for sample-cluster `9'):
    This gene-cluster shows distinct up-regulation pattern for sample cluster `9', and it includes an interesting gene `IL-6'.
IL-6 is known to be a central target gene in a synergistic crosstalk between NF-$\kappa$B and JAK/STAT pathway, which is a unique feature for some DLBCL \cite{Lam2008}. It is suggested that, these have implications for targeted therapies by blocking STAT3 expression, a gene that is activated by IL-6 \cite{Lam2008,Ding2008}.
\item Gene cluster `102' (significant for sample-cluster `11'):
This gene-cluster particularly down-regulates the CLL sampe-cluster among all lymphoma-related clusters. Though it does not indicate a particularly significant GO term (see Table 2 in the supporting material), it includes a number of genes related to regulating tumor cell growth for CLL (See Table 3 in supporting material for the list of genes). Among these genes, let us note IRF1, which is a well-known mediator for cell fate by facilitating apoptosis, and it is also a tumor suppressor \cite{Romeo2002}.
As the expression of IRF1 is slightly down-regulated, we suspect that this may contribute to the growth of CLL tumor cells.
\end{itemize}

In conclusion let us stress that these results strongly indicate that the DBHT technique can detect relevant differentiation and aggregations in both cancer-samples and gene-clones revealing important  relations that can be used for diagnosis, for prognosis and for treatment of these human cancers.

\section*{Discussion}
In summary, we have introduced a novel approach, the DBHT technique,  to extract cluster structure and to detect hierarchical organization in complex data-sets.
This approach is based on the study of the properties of topologically embedded graphs built from a similarity measure.
The DBHT technique is deterministic, it requires no a-priori parameters and it does not need any expert supervision.
We have shown that the DBHT technique can successfully retrieve the clustering and hierarchical structure both from artificial data-sets and from different kinds of real data-sets outperforming in several cases other established methods. The application of the DBHT technique to a referential gene-expression dataset \cite{Alizadeh00} shows that this method can be successfully used in differentiating patients with different cancer subtypes from gene-expression data.
In particular, we have correctly retrieved the differentiation into distinct clusters associated with cancer subtypes (FL, CL and DLBCL)  along with a meaningful hierarchical structure.
The DBHT technique provides a meaningful differentiation of the DLBCL cancer samples into four distinct clusters which turn out to correspond to different survival rates.
The application of the DHBT clustering technique over the gene-clones identifies new groups of genes that play a relevant role in the differentiation of the cancer subtypes, and possibly in relevant genetic pathways which control survival/proliferation of the tumor cells.
Differently from \cite{Alizadeh00} which indicates GCB- and ABC-like DLBCL classification under thorough supervision with biological expertise, we have found instead,  in a completely un-supervised manner, four subtypes of DLBCL with different expression signatures that differentiate significantly in their genetic mechanisms and biological features resulting in well distinct survival rates, hence providing a new perspective. 
It should be stressed that the DBHT technique is addressing the problem of data clustering and hierarchical study from a different perspective with respect to other approaches commonly used in the literature.
It therefore provides an important alternative support in a field where the sensitivity of the results to the kind of approach is often crucial.
The DBHT technique can be extended to more complex measures of dependency which may be also asymmetric.
In our graph theoretic approach this can be handled by constructing topologically embedded directed graphs.
Another extension may concern the use of graph-embedding on  surfaces of genus larger than zero that will provide more complex networks and a richer data filtering \cite{Aste2005}.

\section*{Acknowledgments}
Many thanks Dr. Rohan Williams for helpful discussions and advices on Gene Ontology analysis and COST MP0801 project.

\begin{figure}[!ht]
\centering
\begin{tabular}{c}
\includegraphics[width=0.8\columnwidth]{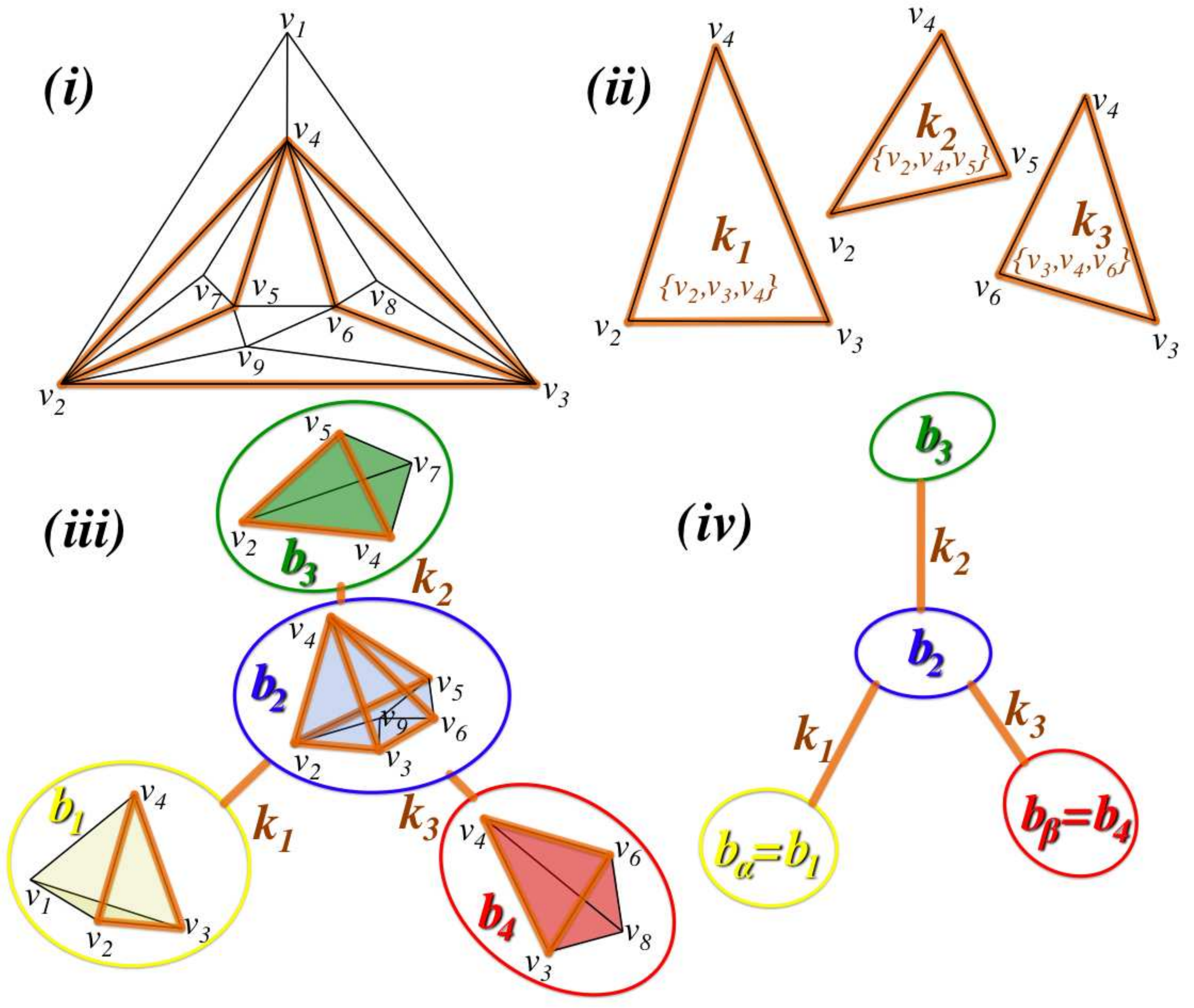}
\end{tabular}
\caption{\label{f.1} A schematic overview of the construction of the bubble tree.
(i)  An example of PMFG graph made of nine vertices $V(G)=\{ v_1,v_2, v_3,v_4,$ $v_5,v_6,v_7,v_8,v_9 \}$ and containing three separating 3-cliques: $k_1$, $k_2$ and $k_3$.
(ii)  The separating 3-cliques have vertex sets: $V(k_1)= \{v_2,v_3,v_4\}$, $V(k_2)= \{v_2,v_4,v_5\}$, and $V(k_3)= \{v_3,v_4,v_6\}$.
(iii) The separating 3-cliques identify four planar sub-graphs called ``bubbles'': $b_1$, $b_2$, $b_3$ and $b_4$ with vertex sets $V(b_1)= \{v_1,v_2,v_3,v_4\}$, $V(b_2)= \{v_2,v_3,v_4,v_5,v_6,v_9\}$, $V(b_3)= \{v_2,v_4,v_5,v_7\}$ and $V(b_4)= \{v_3,v_4,v_6,v_8\}$.
(iv) The graph can be viewed as a ``bubble tree'' made of four bubbles connected through three separating 3-cliques.
}
\end{figure}

\begin{figure}[!ht]
\centering
\begin{tabular}{c}
\includegraphics[width=0.8\columnwidth]{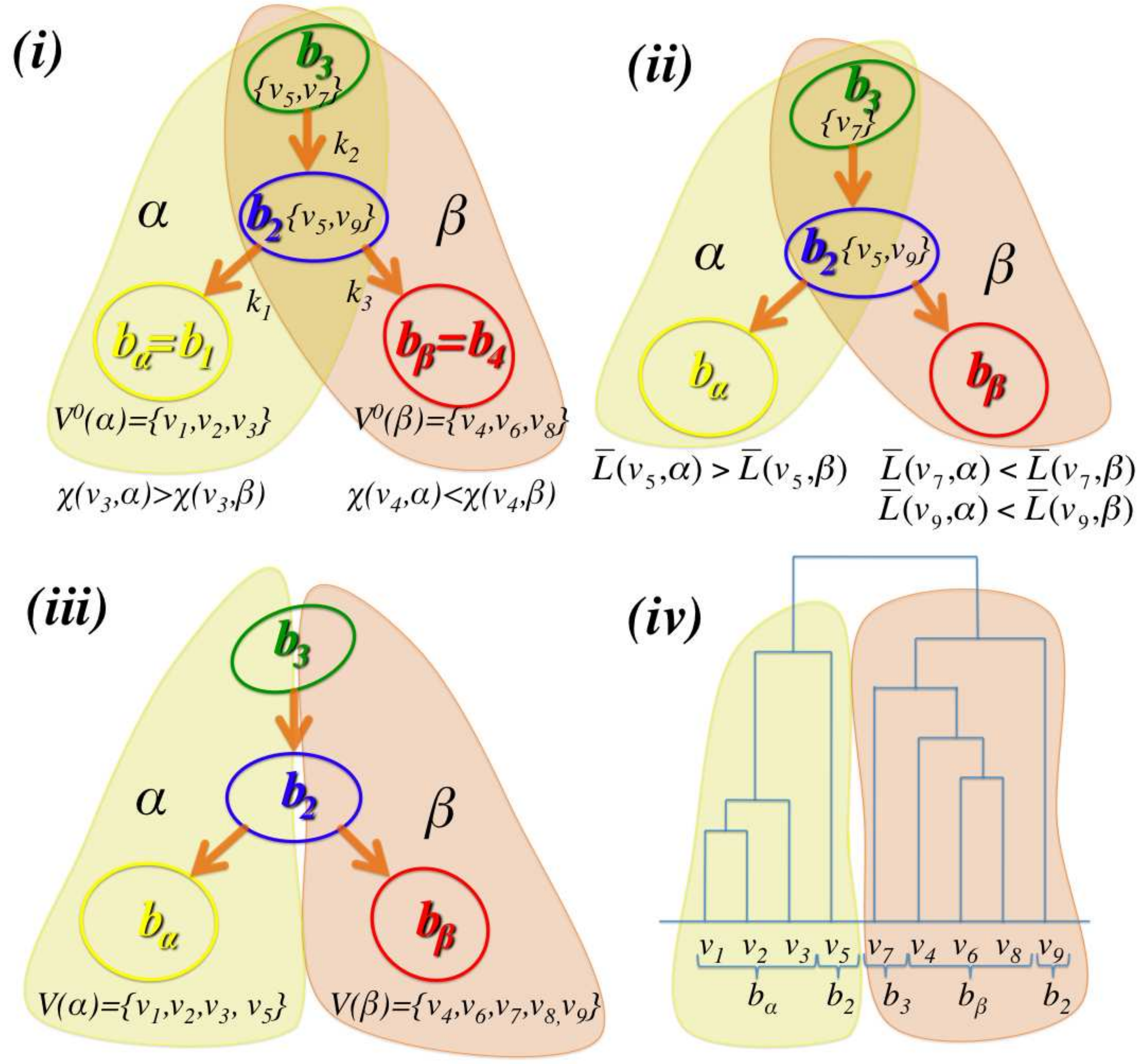}
\end{tabular}
\caption{\label{f.2}
Illustration of the DBHT technique.
(i)  Construction of the directed bubble tree where directions  are given to the 3-cliques $k_1$, $k_2$ and $k_3$ (from Fig.\ref{f.1}) accordingly with the largest weight $W_p^{in}$ and $W_p^{out}$ (Eq.\ref{W} in Methods section).
 In this example we have two converging bubbles: $b_\alpha = b_1$ and $b_\beta = b_4$.
A unique set of vertices can be associated to each of the two converging bubbles $b_\alpha$ and $b_\beta$ where vertices shared by both the converging bubbles (i.e. the vertices $v_3$ and $v_4$) are assigned accordingly with the largest strength $\chi$ (Eq.\ref{strength} in Methods section).
(ii) All the other non-assigned vertices (i.e. $v_5$, $v_9$ and $v_7$) are associated to the cluster with minimum average shortest path length $\bar L$ (Eq.\ref{Lbar} in Methods section).
(iii) The vertex set is uniquely divided into two clusters respectively associated to the two converging bubbles: $V(\alpha) = \{v_1,v_2,v_3,v_5 \}$ and $V(\beta) = \{v_4,v_6,v_7,v_8,v_9 \}$.
(iv) The hierarchical organization and the clustering structure can be represented with a dendrogram.
}
\end{figure}

\section*{Methods}
The PMFG is a weighted graph where edges $uv$ have weights $w_{u,v}$ which, in general, are similarity measures (a larger weight $w_{u,v}$ of edge $uv$ corresponds to a stronger similarity between $u$ and $v$).
Furthermore, a distance $d_{u,v}$, or more generically, a non-negative dissimilarity measure is also associated to the edges.
Specifically, the PMFG is a graph $G(V,E,W,D)$ where $V$ is the vertex set, $E$ the edge set, $W$  the edge-weight set and $D$  the edge-distance set.
A hierarchy in $G$ can be built from a simple consequence of planarity which imposes that any cycle (a closed simple path with the same starting and ending vertex) must be either separating or non-separating  \cite{Diestel2005}.
If we detach from the graph the vertices belonging to a separating cycle then two disjoint and non-empty subgraphs are produced.
The simplest cycle is the 3-clique which is a key structural element in PMFGs.
An example of PMFG is shown in Fig.\ref{f.1} where the separating 3-cliques are highlighted.
By definition, each separating 3-clique, $k_p$, divides the graph $G$ into two disconnected parts, the \emph{interior} $ G_p^{in}$ and the \emph{exterior} $G_p^{ex}$, that are joined by the clique itself. The union of one of these two parts and the separating clique is also a maximally planar graph.
Such a presence of cliques within cliques provides naturally a hierarchy.
The subdivision process can be carried on until all separating 3-cliques in $G$ have been considered.
The result is a set of planar graphs, that we call ``bubbles'', which are connected to each other via separating 3-cliques, forming a tree \cite{Song2011}.
In Fig.\ref{f.1}(a) the ``bubble tree'' and its construction are shown.
In the bubble tree (${H_b}$) vertices $b_i$ represent bubbles and edges $b_i b_j$ represent the separating 3-clique, $k_p$, which is connecting the two bubbles.
A direction can be associated to each edge in ${H_b}$ by comparing the sum over the weights of the edges in the PMFG connecting the 3-clique $k_p$ with the two bubbles.
Specifically, a direction can be associated to the edge $b_i b_j$ by comparing the connections of $k_p$ with the interior sub-graph $ G_p^{in}$ and the exterior sub-graph $G_p^{ex}$ and considering the two weights
\begin{equation}\label{W}
W_p^{in/ex}=\sum_{v\in k_p,u\in G_p^{in/ex}}A_G(v,u)
\end{equation}
where $A_G(v,u)=w_{vu}$ is the adjacency matrix of $G$.
The direction is given toward the side with largest weight obtaining  $\overrightarrow{H_b}$.
(In the case of equal weights in the two directions, the two bubbles are joined into a single larger bubble.)
In $\overrightarrow{H_b}$ there are three different kinds of bubbles:
(1) \emph{ converging bubbles} where the connected edges are all incoming to the bubble;
(2) \emph{ diverging bubbles} where the connected edges are all outgoing from the bubble;
(3) \emph{passage bubbles} where there are both inwards and outwards connected edges.
An example is provided in Fig.\ref{f.2} where we have two converging bubbles ($b_1$ and $b_4$), one diverging bubble ($b_3$) and one passage bubble ($b_2$).
Converging bubbles are special being the end points of a directional path that follows the strongest connections and we consider them as the centers of clusters.
Any bubble $b_i$ connected by a directed path in $\overrightarrow{H_b}$ to a converging bubble $b_\alpha$ belongs to cluster $\alpha$.
By construction, bubbles in cluster $\alpha$ form a subtree $\overrightarrow{h_{\alpha}}$ which has only one converging bubble $b_\alpha$ and all edges are directed toward $b_\alpha$.
This is a non-discrete clustering of bubbles because there can be multiple directed paths between $b_i$ and two or more converging bubbles $b_\alpha$, $b_\beta$,...~.
In Fig.\ref{f.2}(ii) the two subtrees converging toward $b_\alpha=b_1$ and $b_\beta=b_4$ are highlighted, it is clear that in this example bubbles $b_2$ and $b_3$ are shared by the two subtrees.
A non-discrete clustering of the vertex set $V(G)$ can now be obtained by assigning to each vertex $v$ the cluster memberships of the bubbles that contain it.
In order to obtain a \emph{discrete} clustering for $V(G)$ we uniquely assign each vertex to the converging bubble which is at the smallest shortest path distance (see Fig.\ref{f.2} for a schematic overview).
This is achieved in two steps.
\emph{First}, we consider the vertices in the converging bubbles.
Some vertices belong to only one converging bubble and, in this case, they are assigned to it (e.g. in Fig.\ref{f.2} vertices $v_1$ and $v_2$ are assigned to $b_\alpha = b_1$ and  vertices $v_6$, $v_8$ are assigned to $b_\beta = b_4$).
Other vertices instead belong to more than one converging bubble (e.g. vertices $v_3$ and $v_4$ in Fig.\ref{f.2}) and in this case we look at the `strength' of attachment
\begin{equation}
\chi(v,b_\alpha) = \frac{\sum_{u \in V(b_\alpha)} A_G(v,u)}{3(|V(b_\alpha)| -2)} \;\;,
\label{strength}
\end{equation}
and assign each vertex to the bubble with largest strength.
(The notation $|V(b_\alpha)|$ in Eq.\ref{strength} indicates the number of vertices in the vertex set of $b_\alpha$ and $3(|V(b_\alpha)| -2)$ is the number of edges in a bubble.)
After this assignment, each converging bubble $\alpha$ has a unique set of vertices $V^0(\alpha)$.
(There can be converging bubbles with an empty set of vertices and, in this case, there will be no clusters associated to them.)
\emph{Second}, we consider all the other remaining vertices (e.g. vertices $v_5$, $v_7$ and $v_9$ in Fig.\ref{f.2}).
A vertex $v$ may belong to more than one subtree $\overrightarrow{h_{\alpha}}$, $\overrightarrow{h_{\beta}}$... and, in this case, it is assigned to the converging bubble that has the minimum mean average shortest path distance
\begin{equation}
\bar L(v,\alpha) = mean\{ l(v,u) | u \in V^0(\alpha) \wedge v \in V(\overrightarrow{h_{\alpha}} ) \}
\label{Lbar}
\end{equation}
with respect to all other converging bubbles.
Here $l(v,u)$ is the shortest path distance on $G$ from $v$ to $u$ (the smallest sum of distances $d_{r,s}$ over any path between $v$ and $u$).
We have now obtained a discrete partition of the vertex set $V(G)$ into a number of sub-sets $V(\alpha)$,   $V(\beta)$,... each respectively associated to the converging bubbles $b_\alpha$, $b_\beta$,...~.

Once a unique partition of the vertex set into discrete clusters has been obtained, we can investigate how each of these clusters is internally structured and how different clusters gather together into larger aggregate structures.
This can be achieved by building a specifically tailored linkage procedure that builds the hierarchy at three levels.
\begin{itemize}
\item[1.]
\emph{Intra-bubble hierarchy:} we must first assign each vertex $v \in V(\alpha)$  to a bubble $b_i$ in the subtree $\overrightarrow{h_{\alpha}}$.
Vertices in the converging bubbles have been already assigned to the sets $V^0(\alpha)$.
For all remaining vertices, the ones belonging to only one bubble are assigned to such bubble (e.g. vertices $v_7$ and $v_9$ in Fig.\ref{f.2}).
Whereas, vertices that are belonging to more than one bubble (e.g. vertex $v_5$ in Fig.\ref{f.2}) are assigned to the bubble that maximizes the strength $\chi(v,b_i)$ (Eq.\ref{strength}).
In this way for every cluster $\alpha$ and for each bubble $b_i$ in $\overrightarrow{h_{\alpha}}$ we have a unique vertex set $V^\alpha (b_i)$ on which we can now perform a complete linkage procedure \cite{Sorensen1948} by using the shortest path distances $l(u,v)$ as distance matrix.
\item[2.]
\emph{Intra-cluster hierarchy:} we perform a complete linkage procedure between the bubbles in $\overrightarrow{h_{\alpha}}$ by using the distance matrix
\begin{equation}
d^I_\alpha(b_i,b_j) = \max \{ l(u,v)| u \in V^\alpha(b_i) \wedge v \in V^\alpha(b_j) \} \;\;.
\label{dI}
\end{equation}
\item[3. ]
\emph{Inter-cluster hierarchy:}  we perform a complete linkage procedure between the clusters by using the distance matrix
\begin{equation}
d^{II}(\alpha,\beta) = \max \{ l(u,v) | u \in V(\alpha) \wedge v \in V(\beta) \} \;\;.
\label{dII}
\end{equation}
\end{itemize}
With this procedure we obtain a novel linkage that starts from the discrete clusters and at higher level joins the clusters into super-clusters and, instead, at lower level splits the clusters into a hierarchy of bubbles and splits the bubbles into a hierarchy of elements.

\clearpage
\bibliographystyle{plos2009}

\newpage
\begin{appendix}
\section*{Supplementary Information}

\begin{figure}
\centering
\includegraphics[width=0.8\columnwidth]{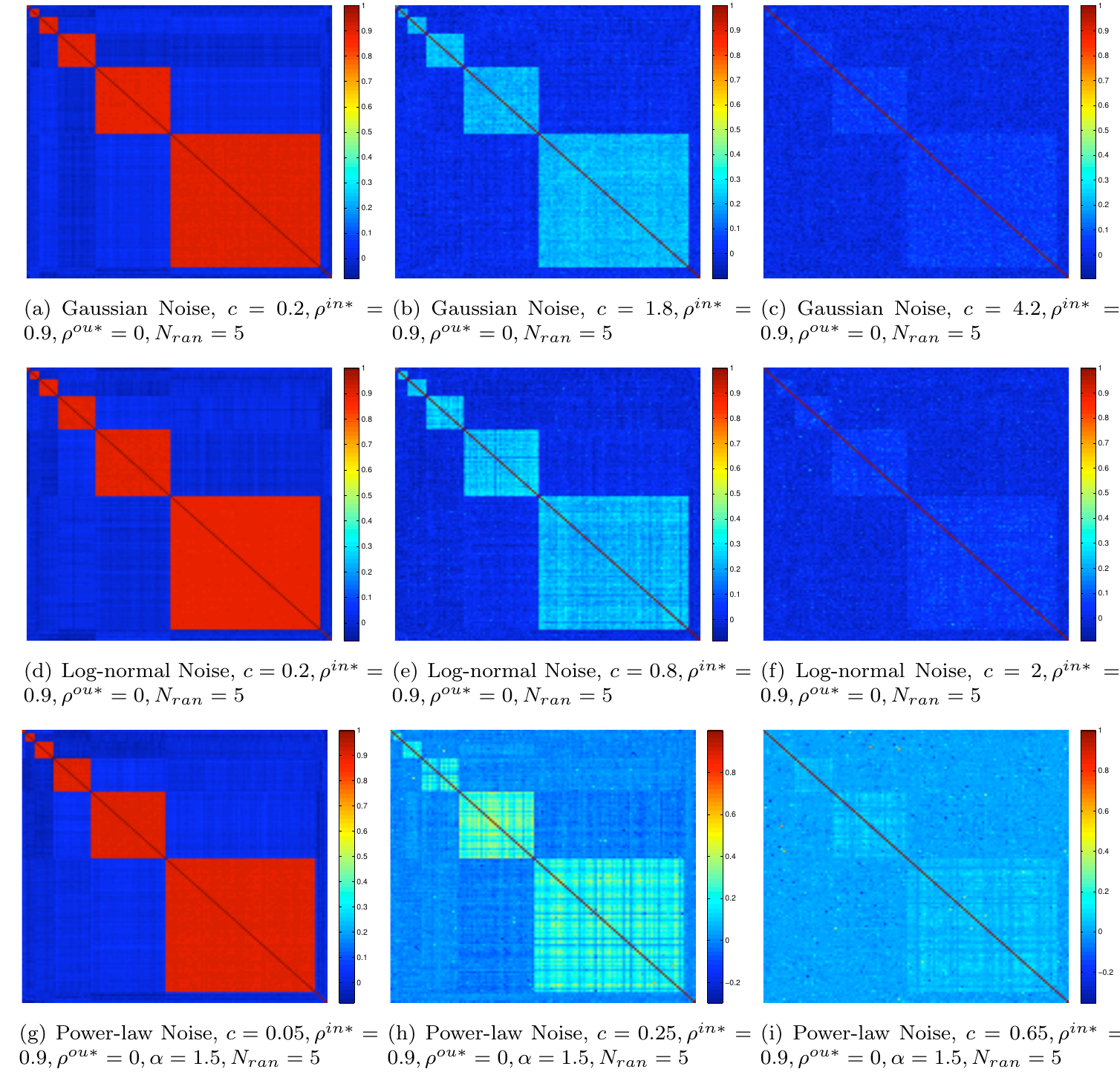}
\caption{\label{R} Visualization of correlation matrices of synthetic data sets generated from MVG with partition of cluster sizes 4,8,16,32 and 64 where relative noise amplitude $c$ has been varied to change the resolution of clustering structure. The parameters are specified underneath each figure. The first row adjusts $c$ for Gaussian noises, the second adjusts for log-normal noises, and the third adjusts for Power-law noises with $\alpha=1.5$.}
\end{figure}

\begin{figure}
\centering
\includegraphics[width=0.8\columnwidth]{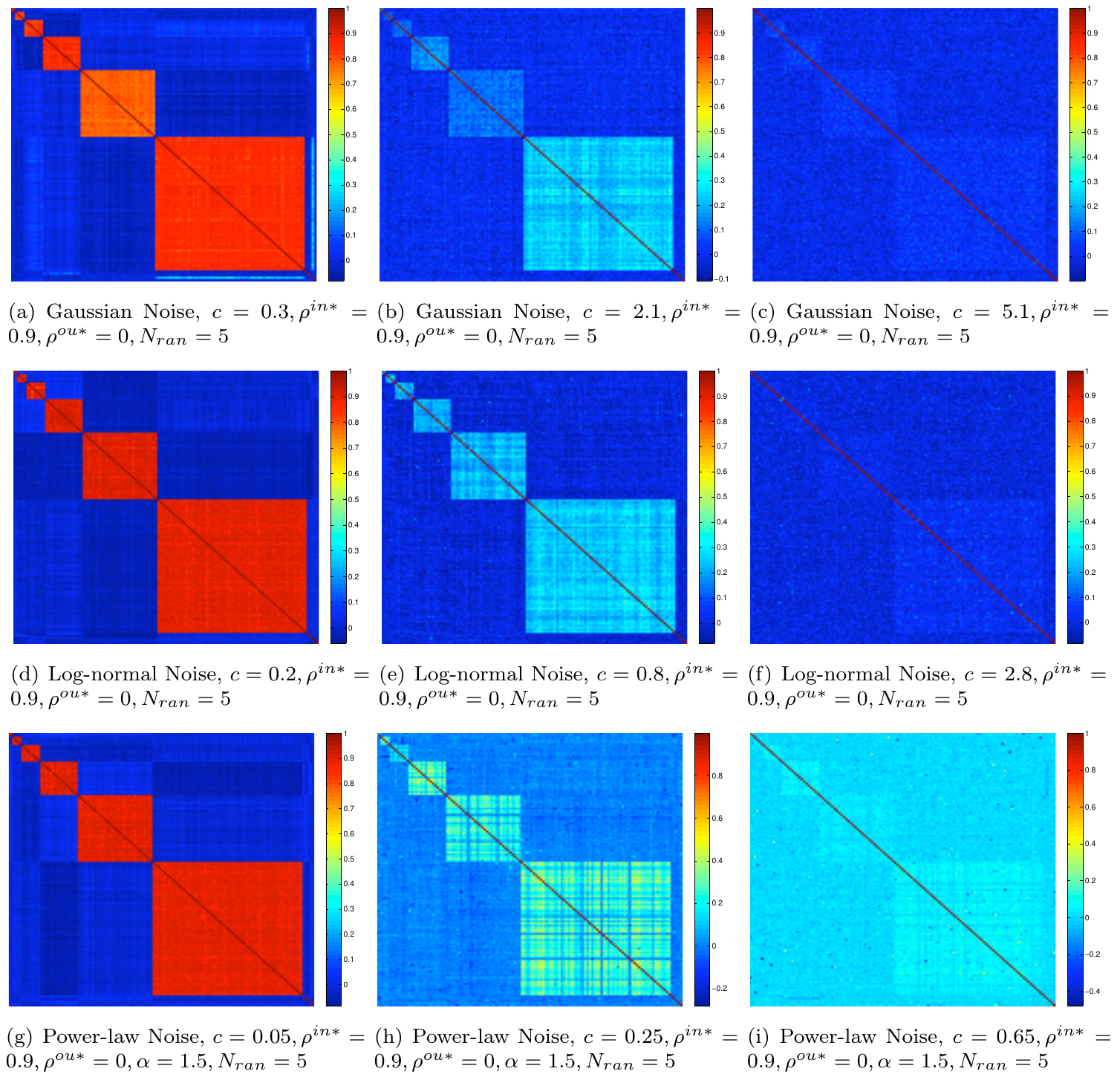}
\caption{\label{LnR} Visualization of correlation matrices of synthetic data sets generated from log-normal multivariates with partition of cluster sizes 4,8,16,32 and 64 where relative noise amplitude $c$ has been varied to change the resolution of clustering structure. The parameters are specified underneath each figure. The first row adjusts $c$ for Gaussian noises, the second adjusts for log-normal noises, and the third adjusts for Power-law noises with $\alpha=1.5$.}
\end{figure}

\section{Artificial data with a clustering structure}
\subsection{Preparation}
By using a multivariate Gaussian generator (MVG) and a multivariate Log-Normal generator[see Wang SS (2004) Casualty actuarial society proc. LXXXV] we have produced several synthetic time series which approximate a given correlation structure $R^*$.
Specifically, we have generated $N$ stochastic time series $y_i(t)$ of length $T$ ($i=1...N$, $t=1...T$) with zero mean and Pearson's cross-correlation matrix $R$ that approximates $R^*$.
As for the starting correlation structures  $R^*$,  we have used block diagonal matrices where the blocks are the artificial correlated clusters.
The matrix $R^*$ has zero inter-cluster correlations $\rho^{ou*}$ and large intra-cluster correlations $\rho^{in*}$ within the diagonal blocks.
To this pre-defined cluster structure, we added a number $N_{ran}$ of random correlations unrelated to the clusters.
We have chosen $T=10\times N$ and  we added a noise term  $\eta_i(t)$ obtaining a new set of dataseries
\begin{equation}
y'_i(t) = y_i(t) +  c \sigma_i \eta_i(t)\;\;,
\end{equation}
where $\sigma_i = \sqrt{\left< y_i^2 \right>-\left< y_i \right>^2}$ is the standard deviation of $y_i(t)$ and $c$ is a constant used to tune the relative amplitude of noise.
We have used a Normally distributed noise with probability distribution function $p(\eta) \propto \exp(-\eta^2/2)$ and  a log-Normally distributed noise with probability distribution function $p(\eta) \propto \exp(-\log(\eta)^2/2)$.
We have varied the relative amplitude of noise $c$ from  0  to  7  with constant intra-cluster correlation in  $R^*$ at $\rho^{in*}=0.9$.
We also have used  power-law distributed noise, with probability distribution function  $p(\eta) \propto 1/\eta^{\alpha+1}$.
Specifically, this noise was numerically generated by using $\eta(t) =  \pm |\eta^{un}(t)|^{(-1/\alpha)}$, where $\eta^{un}(t)$ is a uniformly distributed noise in $(0,1]$ and the sign in front is chosen at random for each $t$ with probability 50\%.
In this case, we have varied the relative amplitude of noise $c$  from 0 to 0.8  with exponent $\alpha = 1.5$ and constant intra-cluster correlation $\rho^{in*}=0.9$.
We also have varied the exponent $\alpha$ between  1  to  3 keeping    $c=0.1$ and  $\rho^{in*}=0.9$.
Examples of the obtained correlation matrices are reported in Fig.\ref{R} for the MVG and Fig.~\ref{LnR} Log-normal multivariate generator.

All these different manipulations produce a similar effect where by increasing the amplitude of noise or by decreasing the exponent or by reducing $\rho^{in*}$,  the Pearson's cross-correlation matrix $R$ passes from a very well defined structure close to $R^*$  to a blurred structure where the average intra-cluster correlation ($\left<\rho^{in}\right>$) becomes smaller and finally it becomes equal to the average inter-cluster correlation ($\left< \rho^{ou}\right> $) and no correlation structure can be any longer observed.

In summary, the simulated data were generated by combining the following possibilities.
\begin{itemize}
\item Partitions:
\begin{itemize}
\item Regular Partitions (all clusters of the same size),
\item Irregular partitions (clusters with different sizes).
\end{itemize}
\item Type of multivariate random variables:
\begin{itemize}
\item Multivariate Gaussian Distribution;
\item Multivariate Log-normal Distribution.
\end{itemize}
\item Type of perturbation noises:
\begin{itemize}
\item Univariate Gaussian Distribution;
\item Univariate Log-normal Distribution;
\item Univariate Power-law Distribution.
\end{itemize}
\item Relative noise amplitude $c$.
\item Random background elements $N_{ran}$.
\end{itemize}

\subsection{Comparison with different clustering methods}
\begin{figure}[ht!]
\centering
\includegraphics[width=0.8\columnwidth]{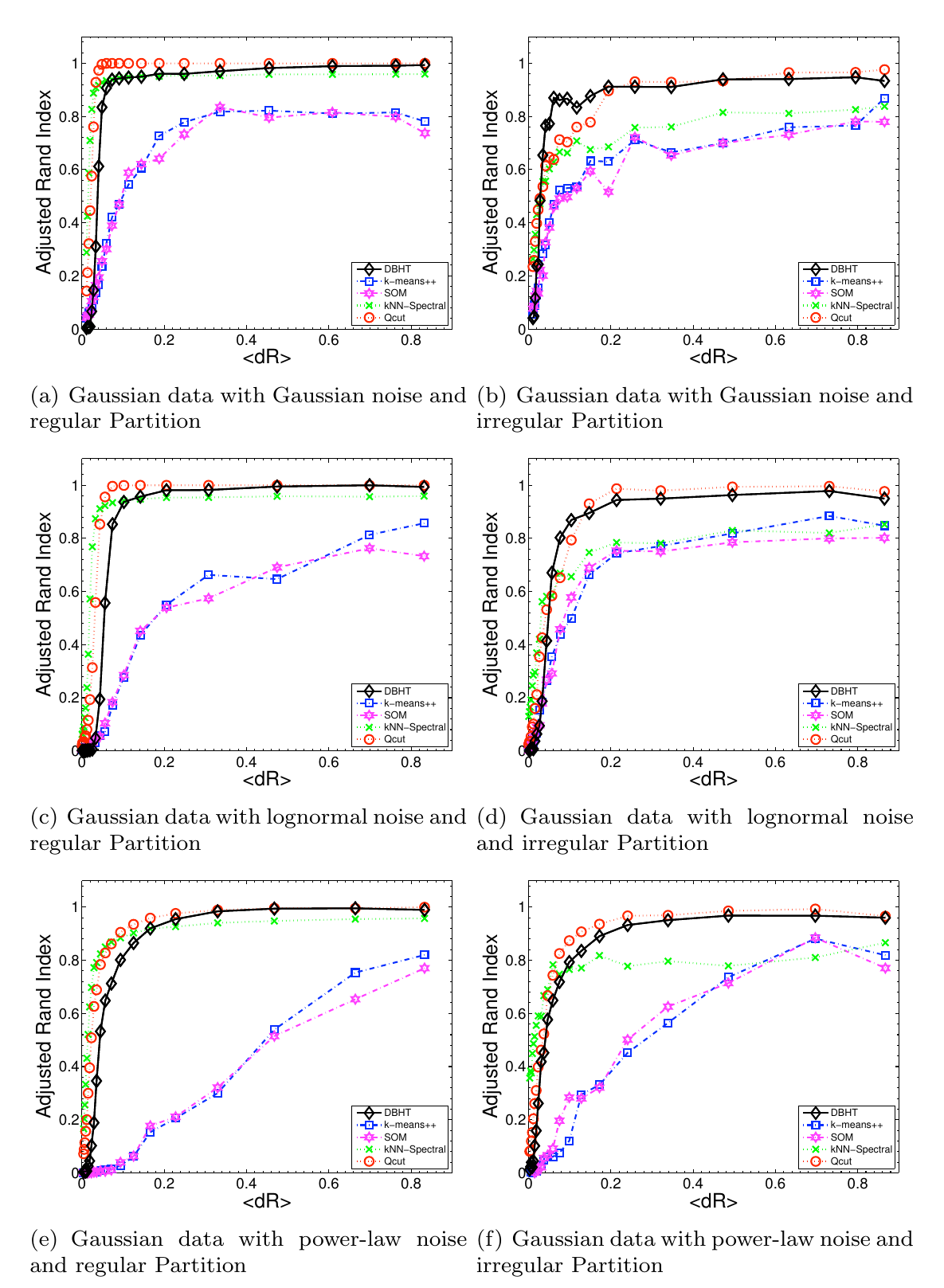}
\label{GaussDataPowerlawNoiseIrregularPartition}
\caption{\label{GaussData}
Adjusted Rand index for various data sets simulated via Gaussian (Normal) distribution with $\rho^{in*}=0.9,\rho^{ou*}=0$ and $N_{ran}=5$.
For each value of $C$, 30 data sets were generated in order to get stable statistics for $<dR>$ and adjusted Rand score.}
\end{figure}

\begin{figure}[ht!]
\centering
\includegraphics[width=0.8\columnwidth]{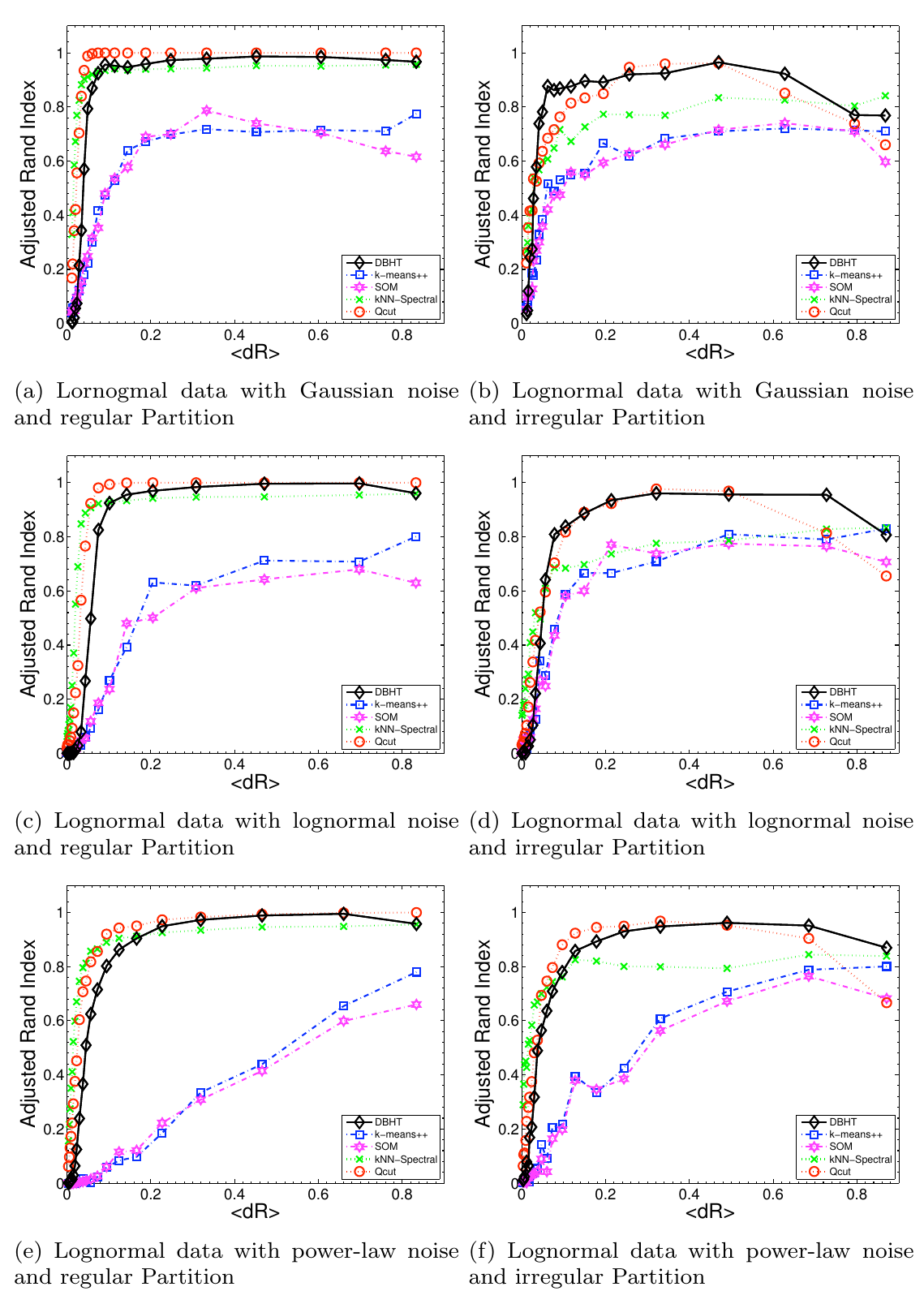}
\label{LognormalDataPowerlawNoiseIrregularPartition}
\caption{\label{LognormalData}
Adjusted Rand index for various data sets simulated via Log-normal distribution with $\rho^{in*}=0.9,\rho^{ou*}=0$ and $N_{ran}=5$.
For each value of $C$, 30 data sets were generated in order to get stable statistics for $<dR>$ and adjusted Rand score.}
\end{figure}

\begin{figure}[ht!]
\centering
\includegraphics[width=0.8\columnwidth]{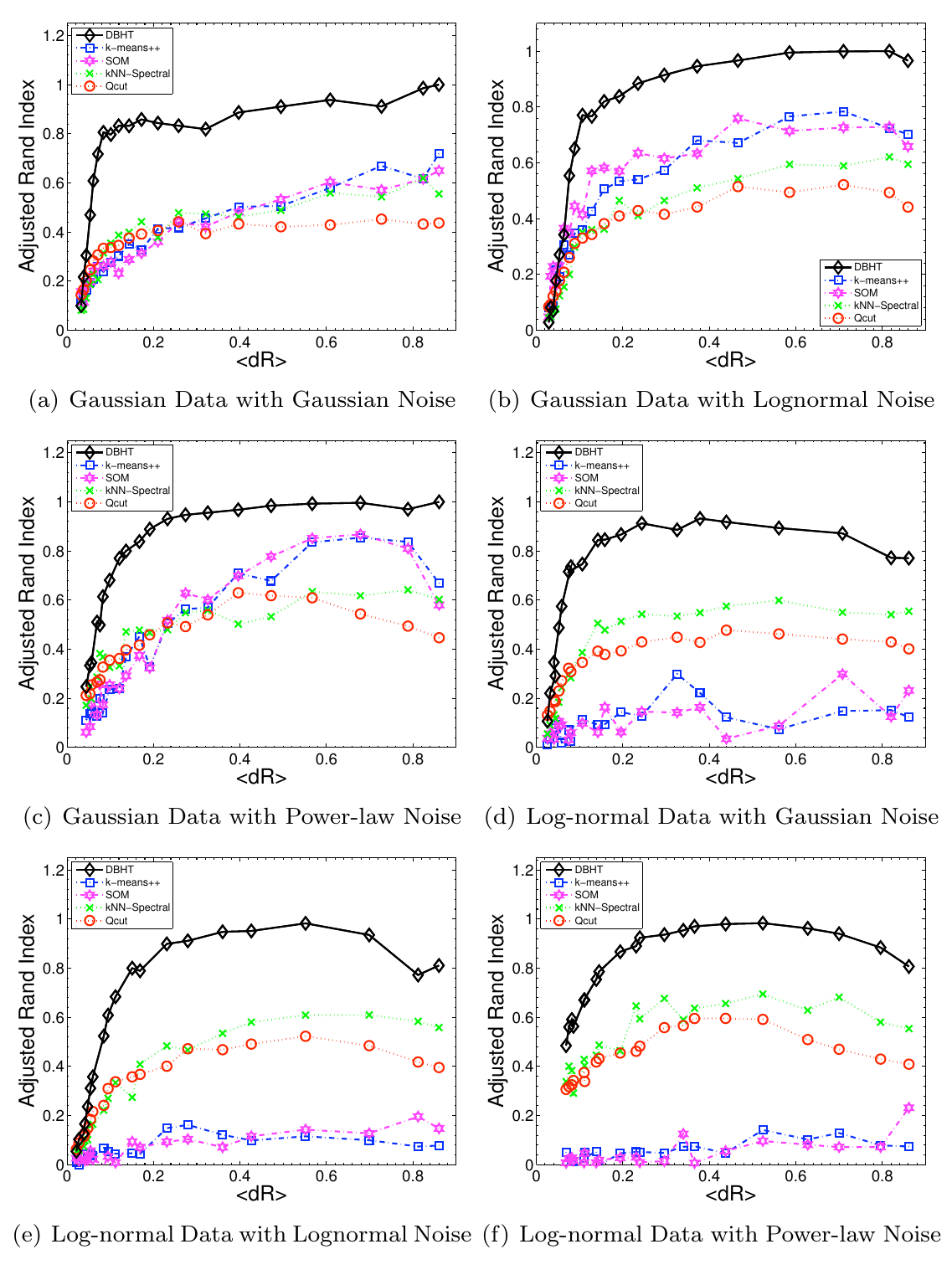}
\label{CaseSpecificLognormalDataPowerlawNoise}
\caption{\label{CaseSpecific}
Adjusted Rand index for various data sets simulated via Gaussian and Log-normal distribution with $\rho^{in*}=0.9,\rho^{ou*}=0$ and $N_{ran}=25$.
This case refers to a cluster structure with eight clusters of size 5 elements, and one cluster of size 64 elements.
For each value of $C$, 30 data sets were generated in order to get stable statistics for $<dR>$ and adjusted Rand score.
Figure (a) and (f) are the same of Fig.1 in the paper and are here reported for completeness and for an easier comparison.
}
\end{figure}

Fig.~\ref{GaussData} shows the performance curves evaluated via adjusted Rand index for simulated data with multivariate Gaussian distribution, and Fig.~\ref{LognormalData} shows the performance curves for simulated data with multivariate Log-normal distribution.
The results for a wide range of $dR>0.1$ for a broad set of combinations  show that DBHT clustering  outperforms the other clustering techniques except for Qcut which performs similarly to the DBHT.
However, Fig.~\ref{CaseSpecific} shows that the DBHT clustering can outperform also Qcut for both Gaussian and Log-normally simulated data when an extreme cluster size differentiation is present.
Specifically,  in Fig.~\ref{CaseSpecific}, there is a structure of eight small clusters of size 5 elements and one big cluster of 64 elements, and large number of random background elements ($N_{ran}=25$).
Let us stress that the performance curves in Fig.~\ref{CaseSpecific} demonstrate that DBHT clustering is the only technique which delivers consistent and quality clustering outcomes in spite of the severe conditions applied.

\section{Artificial data with a hierarchical Structure}
\subsection{Preparation}
In order to test the DBHT technique for the detection of the hierarchical structure, we have generated input matrices $R^*$ that are organized in a nested block-diagonal structure where block of small sizes are placed inside blocks of lager sizes.
In particular, we looked at regular partitions of 16 `small' clusters containing 16 elements each with $\rho^{in*}_1=0.95$.
These small clusters are merged to `medium' clusters with $\rho^{in*}_2=0.8$, and further merged to `big' clusters with $\rho^{in*}_2=0.7$.
Finally, all clusters are merged to a single cluster with $\rho^{ou*}=0.15$.
Similarly, we looked at irregular partitions with clusters of scaling sizes containing, 4, 4, 8, 8, 16, 16, 32, 32, 64 and 64 elements each, and the structures of small, medium, and big clusters were embedded by consecutively merging with $\rho^{in*}_1,\rho^{in*}_2,\rho^{in*}_3$ and $\rho^{ou*}$.

\begin{figure}[t]
\centering
\includegraphics[width=0.8\columnwidth]{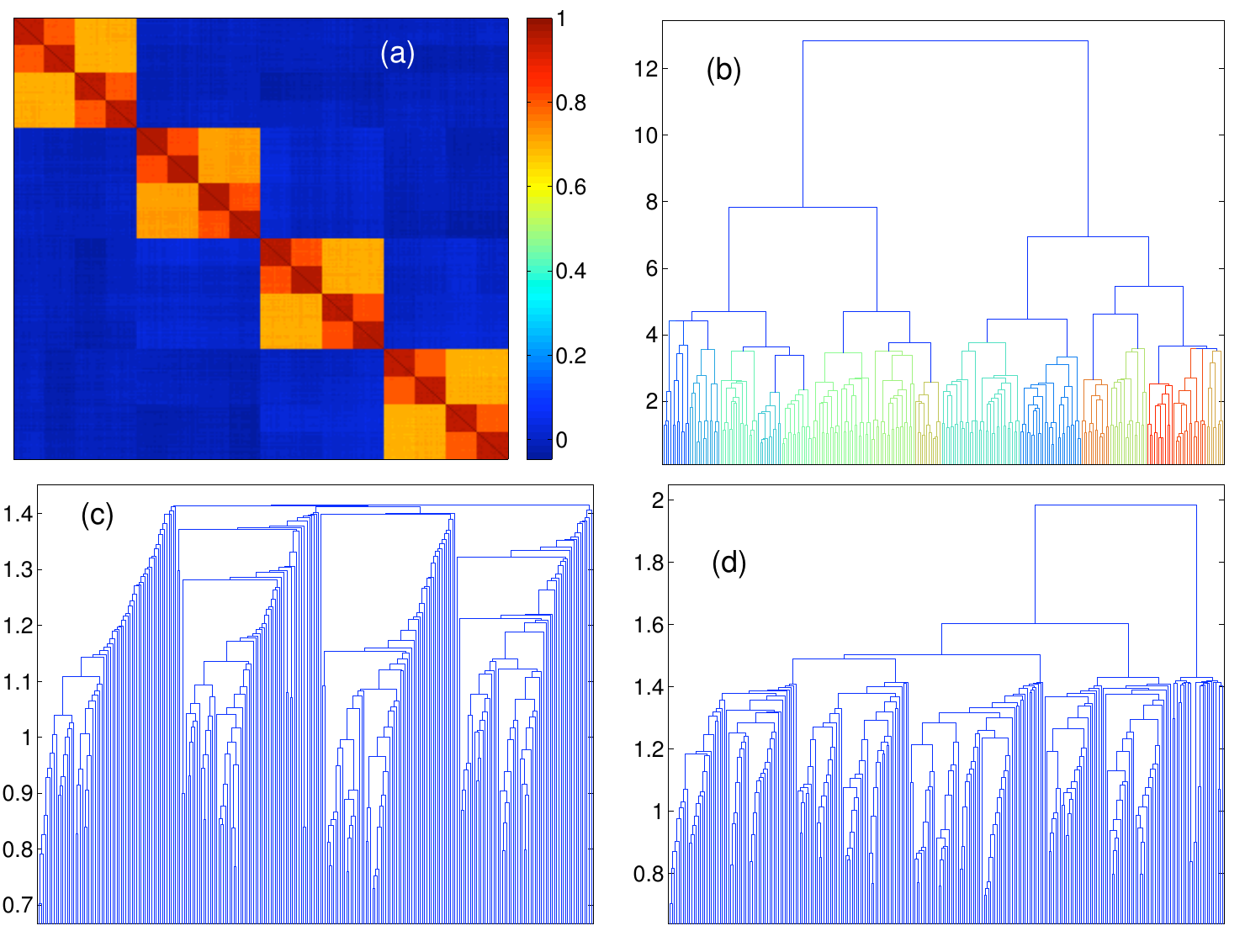}
\caption{\scriptsize \label{f.4xS1}
Hierarchical clustering for uniform partition with a power law noise with exponent $\alpha =1.1$ and noise level $c=0.03$
{\bf (a)} Correlation template $R^*$ for a synthetic data structure  with uniform sizes of 16 elements each.
{\bf (b)} Dendrogram associated with the DBHT hierarchical structure.
{\bf (c)} Dendrogram associated with the Average linkage.
{\bf (d)} Dendrogram associated with the Complete linkage.
}
\end{figure}

\begin{figure}[tb]
\centering
\includegraphics[width=0.8\columnwidth]{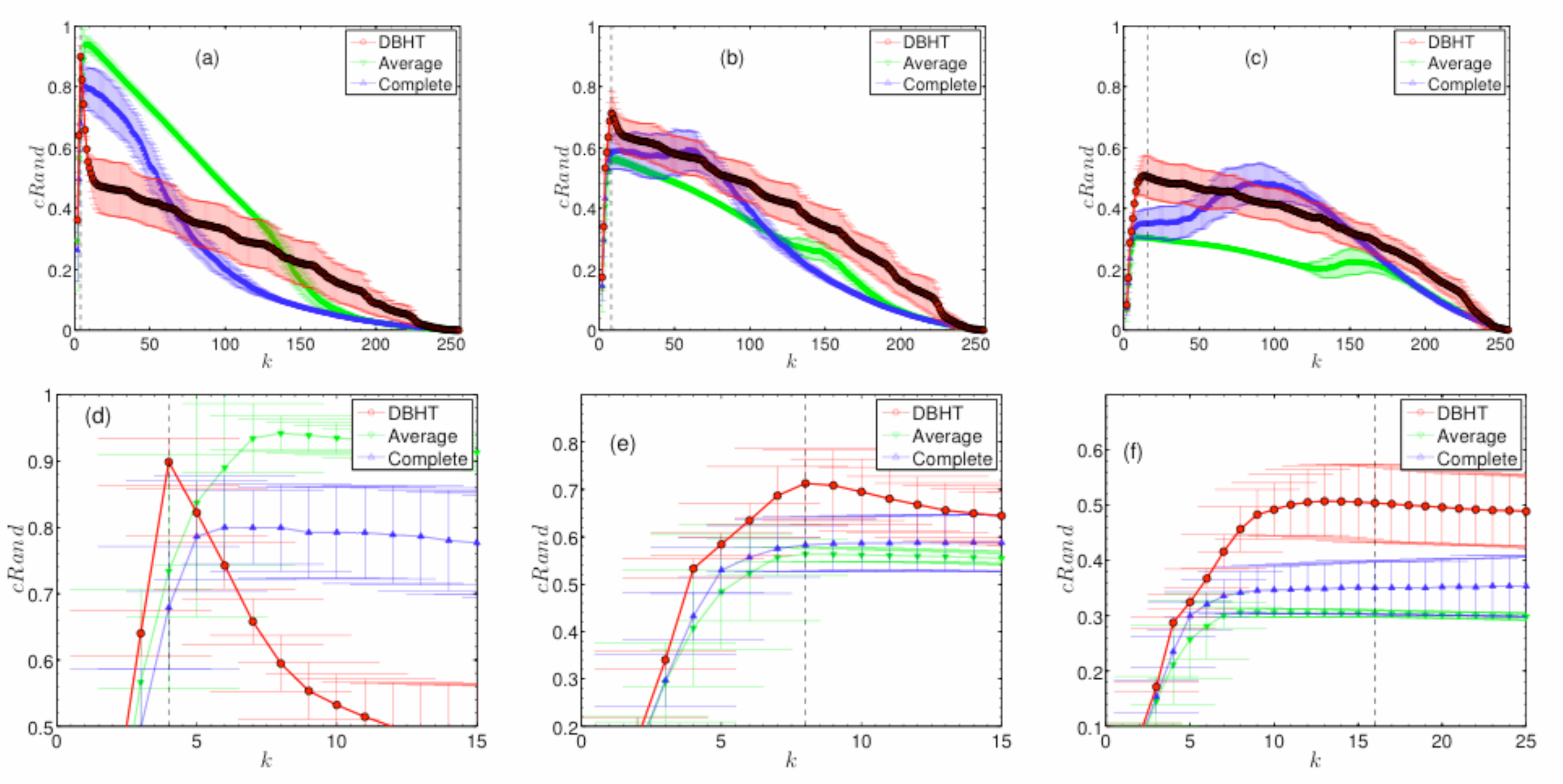}
\caption{\scriptsize \label{f.4hierS1}
Adjusted Rand index for the comparison between the synthetic partition in Fig.\ref{f.4xS1}(a) and the partitions retrieved by cutting the dendrograms from our DBHT clustering method at various numbers of clusters.
{\bf (a)} Comparison between the  synthetic partition with the 4 large clusters and the partitions from DBHT, average linkage and complete linkage.
{\bf (b)} Comparison between the  synthetic partition with the 8 medium clusters and the partitions from DBHT, average linkage and complete linkage.
{\bf (c)} Comparison between the  synthetic partition with the 16 small clusters and the partitions from DBHT, average linkage and complete linkage.
{\bf (d),(e),(f)} Details of the upper figures showing the region where the DBHT has the maximum.
The plots report average values over a set of the 30 trials, the error bars are the standard deviations.
}
\end{figure}

\begin{figure}[t]
\centering
\includegraphics[width=0.8\columnwidth]{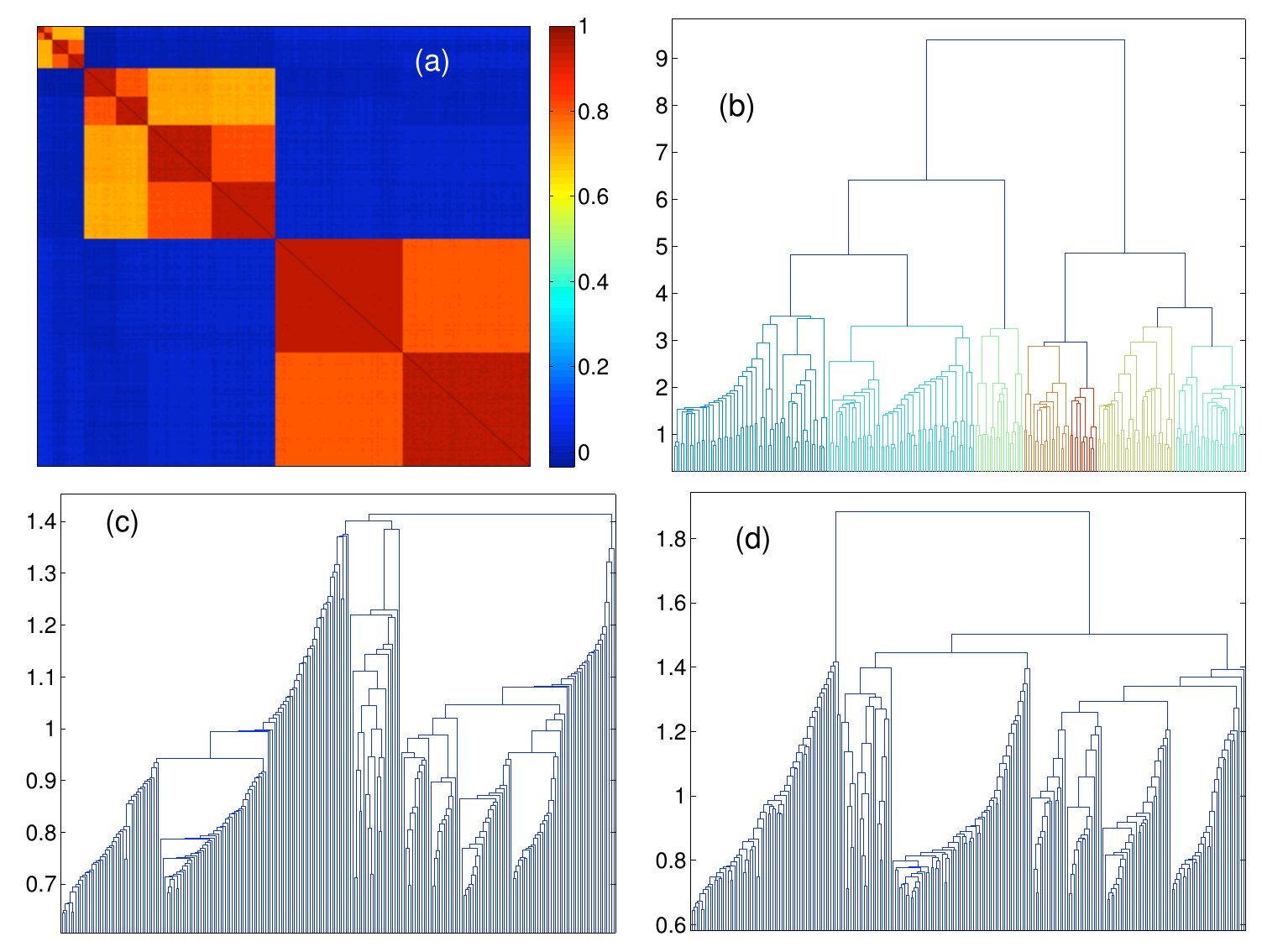}
\caption{\scriptsize \label{f.4xS}
{\bf (a)} Correlation template $R^*$ for a synthetic data structure with clusters with scaling sizes of 4, 8, 16, 32 and 64.
{\bf (b)} Dendrogram associated with the DBHT hierarchical structure.
{\bf (c)} Dendrogram associated with the Average linkage.
{\bf (d)} Dendrogram associated with the Complete linkage.
}
\end{figure}

\begin{figure}[tb]
\centering
\includegraphics[width=0.8\columnwidth]{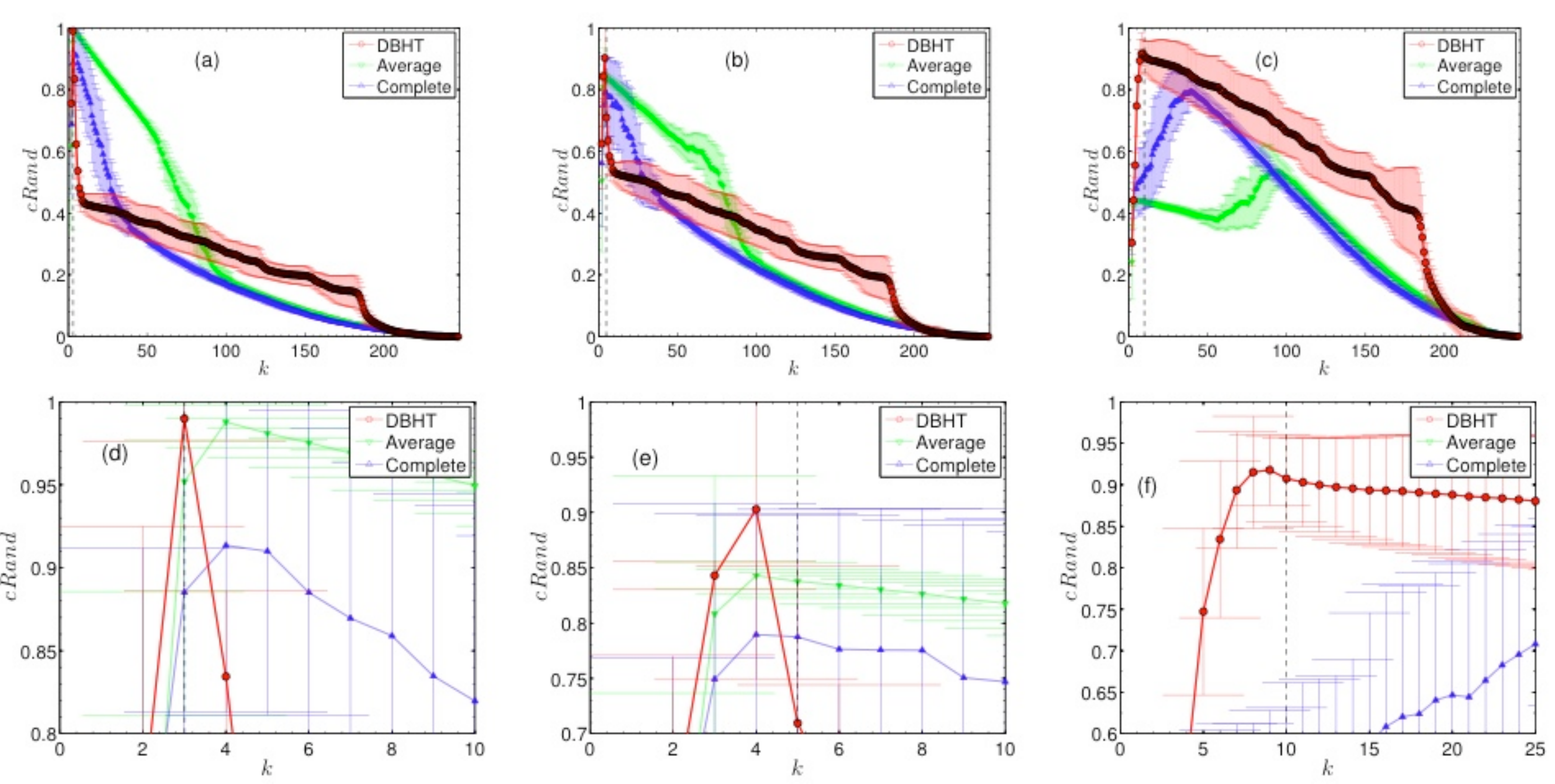}
\caption{\scriptsize \label{f.4hierS}
Adjusted Rand index for the comparison between the synthetic partition in Fig.\ref{f.4xS}(a) and the partitions retrieved by cutting the dendrogram from the DBHT clustering method at various number of clusters.
{\bf (a)} Comparison between DBHT clustering and the synthetic partition with the 2 `large' clusters.
{\bf (b)} Comparison between DBHT clustering and the synthetic partition with the 5 `medium' clusters.
{\bf (c)} Comparison between DBHT clustering and the synthetic partition with the 10 `small' clusters.
{\bf (d),(e),(f)} Details of the upper figures showing the region where the adjusted Rand index from DBHT has the maximum.
The plots (b), (c) and (d) report average values over a set of the 30 trials, the error bars are the standard deviations.
}
\end{figure}

\subsection{Comparison with different linkage methods}
We have simulated 30 different sets of multivariate Gaussian data series of length $T=10\times N$ by using nested hierarchical block-diadonal input matices $R^*$.
An example of $R^*$ is provided in Fig.\ref{f.4xS1}(a) (same as Fig.2(a) in the paper).
We have tested the capability of the DBHT method to recognize hierarchies by moving through the different hierarchical levels varying the number of clusters from only one at the top hierarchy to the number of elements at the lowest hierarchy.
At each number of clusters we have measured the adjusted Rand index with respect to the `large', `medium' and `small' partitions.
Figs.\ref{f.4hierS1}(b-d) show the average adjusted Rand index and the standard deviations over the 30 sets of synthetic data obtained by using the DBHT method, the average linkage method and the complete linkage method.
One can observe in Fig.\ref{f.4hierS1}(b) that all three methods successfully detect the 4 large clusters retrieving adjusted Rand index near to unity.
At following hierarchical levels only the DBHT method consistently retrieves the maximum value for the adjusted Rand index respectively at the hierarchical partitions with 8 and 16 clusters.
Conversely, the other two methods achieve lower maximal values of the adjusted Rand index at a larger number of clusters inconsistent with the sizes of the synthetic data structure.
We have tested other partitions and different levels of noise  verifying that the DBHT method is consistently delivering good performances in comparison with the other established methods.
An example, by using power law noise and clusters of scaling sizes respectively of 4, 8, 16, 32 and 64 elements is reported in Fig.\ref{f.4xS}(a).
The dendrograms for the DBHT, and the average linkage and the complete linkage methods are respectively reported in Figs.\ref{f.4xS}(b,c,d).
The comparison between the  adjusted  Rand indexes is reported in Fig.\ref{f.4hierS}.
One can see that, also in this case, the DBHT technique consistently outperforms the linkage methods.

\section{Lymphoma data analysis}

\subsection{Emergence of GCB-like and ABC-like Patterns on PMFG}

\begin{figure}[tb]
\centering
\includegraphics[width=0.8\columnwidth]{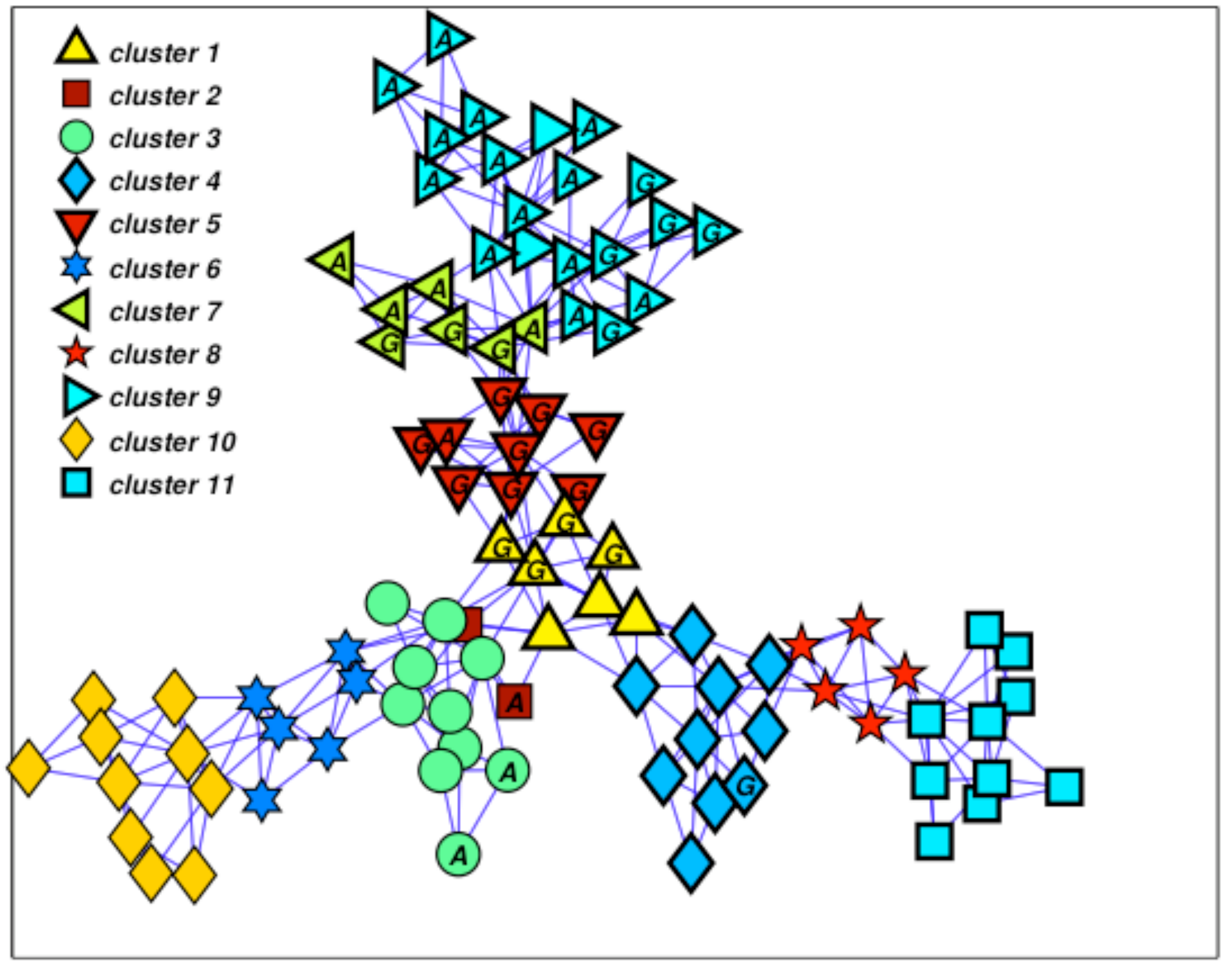}
\caption{Visualization on the PMFG of the GCB-like and ABC-like classifications as given by Alizadeh \emph{et al} 2000.
The labels inside the symbols correspond respectively to GCB-like DLBCL (G) and ABC-like DLBCL (A).
The symbols are the same used to represent the sample clusters found by DBHT technique in Fig. 4 in the main paper \label{GCB_PMFG}.
}
\end{figure}

Here, we report how the GCB-like and ABC-like classification of DLBCL subtypes naturally emerges in the PMFG.
This is shown in Fig.~\ref{GCB_PMFG} where we can observe that ABC-like DLBCLs are dominant on the top of PMFG, and mainly occupy sample-cluster `7' and `9'.
On the other hand, GCB-like DLBCLs are dominant on the center of PMFG, and mainly occupy sample-cluster `1', `5' and `7'. Among the sample-clusters associated with DLBCL, sample-cluster `1' and `5' are distinctively characterized by GCB-like DLBCL, sample cluster `9' is characterized by ABC-like DLBCL. Interestingly, sample cluster `1' and `5' indicate a further sub-classification of GCB-like DLBCL, and yet show superior survival rates than sample clusters associated ABC-like DLBCL, a more fatal subtype indicated by Alizadeh \emph{et al} 2000 than GCB-like DLBCL (See Table 1 in the main paper).
Furthermore, sample-cluster `7' is a mixture of these two subtypes, and yet shows much worse survival rates than sample-cluster `9' in which is present a much larger portion of ABC-like DLBCL (See Table 1 in the main paper).
This clearly shows that the DBHT clustering indicates further meaningful subtypes of DLBCLs with respect to the GCB-/ABC-like classification of Alizadeh \emph{et al} 2000.

\subsection{Analysis of significant gene-clusters for sample-clusters }
\begin{figure}[ht!]
\centering
\includegraphics[width=0.8\columnwidth]{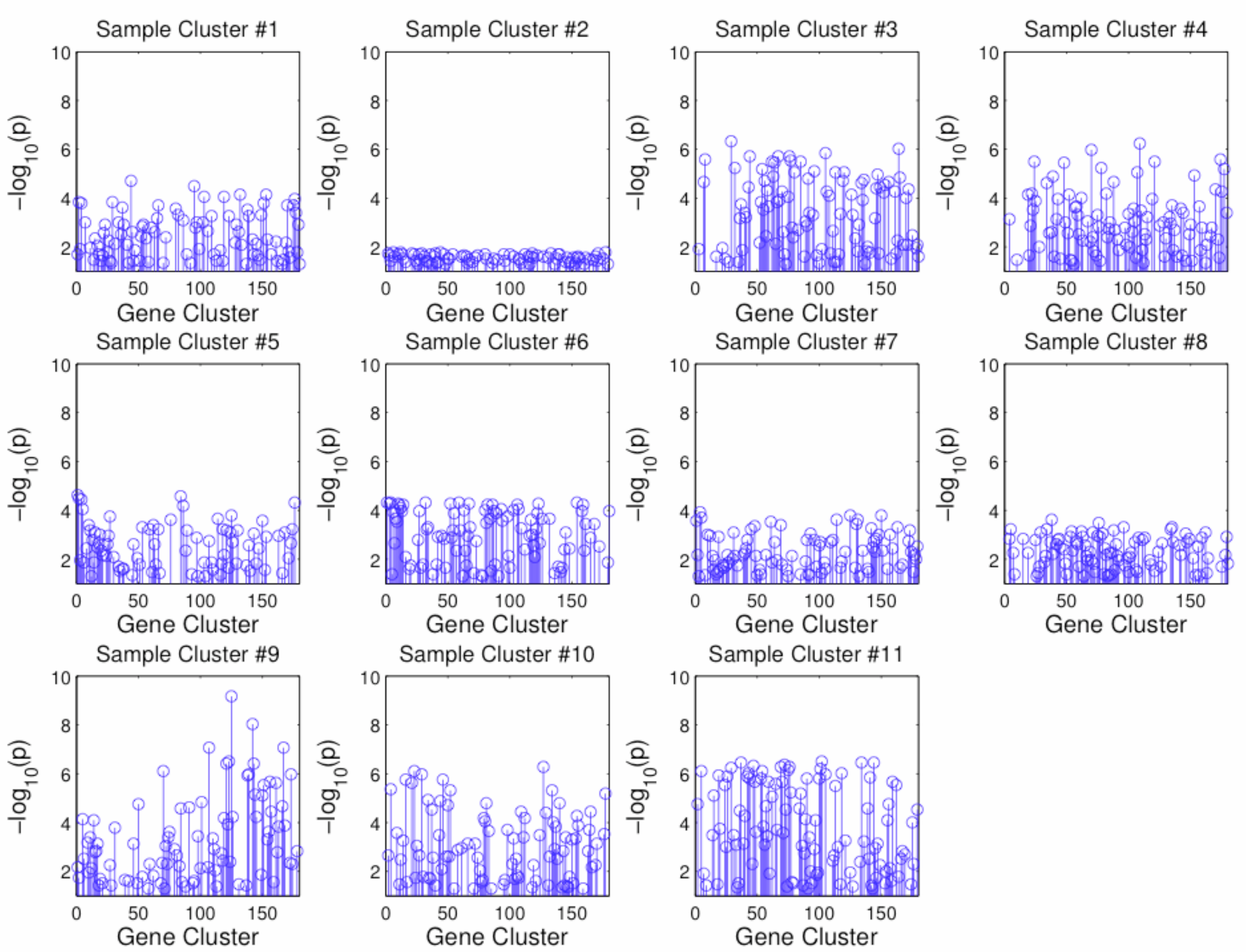}
\caption{\label{CountEnriched}Plot of cut-off p-value for Kruskal-Wallis one-way ANOVA test -vs- enriched gene-clusters. Circles represent the smallest cut-off p-value for individual gene-clusters.}
\end{figure}

In order to look for significant gene-clusters which distinguish each sample-cluster, we have performed a series of statistical analysis on the gene-clusters of the data found by DBHT clustering.
Specifically, we have performed a combination of differential expression and enrichment analysis.
Firstly, for a given sample-cluster, we have looked for a set of differentially expressed gene-profiles for a given cut-off p-value.
Then we have calculated enrichment statistics for each gene-cluster by asking whether this cluster significantly enriches for the differentially expressed profiles.
By varying the cut-off p-values, we have identified the most significant gene-cluster for the particular sample-cluster by choosing the gene-cluster that remains significantly enriched for the smallest cut-off.
In order to identify differentially expressed profiles for each cut-off p-value, we have performed non-parametric Kruskal-Wallis one-way ANOVA test.
The enrichment statistics has been evaluated by using the hypergeometric test with significance level of p-value 0.05, where the p-values were adjusted by Bonferroni correction.
Fig.~\ref{CountEnriched} reports the smallest cut-off p-values for each gene-cluster, for each sample-cluster.
The list of labels for the most significant gene-clusters is shown in Table~\ref{EnrichedGeneClusterList}.
Except for sample-cluster `2' and `6', each sample-cluster is assigned to a unique gene-cluster.
For what concerns  sample-cluster `2' this is most likely due to the small cluster size.
Instead, we note that sample-cluster `6' corresponds to a collection of T Cell samples, and we suspect that the emergence of multiple significant gene-clusters is due to the broad spectrum of T cells in the physiology of lymphoma.
\begin{table}
\centering
\begin{tabular}{|c|c|}			
\hline\hline			
Sample Cluster	&	Gene Cluster	\\
\hline
\emph{\textbf{1}}	&	44	\\
2	&	6,12,44,177	\\
3	&	29	\\
\emph{\textbf{4}}	&	109	\\
\emph{\textbf{5}}	&	1	\\
6	&	1,4,32,59,154	\\
\emph{\textbf{7}}	&	4	\\
8	&	38	\\
\emph{\textbf{9}}	&	125	\\
10	&	127	\\
\emph{\textbf{11}}	&	102	\\
\hline\hline			
\end{tabular}			
\caption{\label{EnrichedGeneClusterList}List of most significant gene-clusters for the sample-clusters. Sample clusters in bold italic font correspond to the clusters associated to lymphoma malignancies.}			
\end{table}

\subsection{Gene Ontology analysis on significant gene-clusters}
Among all significant gene-clusters, we have chosen a subset of gene-clusters which are associated to lymphoma malignancies, and we have performed Gene Ontology (GO) analysis on these gene-clusters in order to investigate associated biological processes.
The analysis has been performed with significance level of p-value 0.05 on a plug-in software for Cytoscape, called BiNGO, and we applied Bonferroni correction.
We have obtained a number of significant biological processes which are reported in Table~\ref{AllSampleClusterGO}.
These biological processes indicate the underlying genetic mechanisms of which genes in the same gene-cluster share.
For instance, gene-cluster `44' is associated to a large number of GO terms for cell cycles and cell cycle regulation.
Indeed, this gene-cluster contains, for various phases, a key cell-cycle regulator CDK1 whose over-expression pattern is a characteristic feature of DLBCL as discussed in the main paper.
On the other hand, none of the significant biological processes was captured by GO analysis for gene-cluster `102'.
However, by no means, this cluster is un-significant for the sample-cluster.
Indeed, as the enrichment analysis in Fig.~\ref{CountEnriched} suggests, gene-cluster `102' remained enriched for very low p-value $\sim 10^{-6}$, and it includes biologically significant genes for CLL such as IRF1 as reported in the main paper.
In Table~\ref{GeneCluster11List} we report the full list of clones for the gene-cluster.

\begin{table}									
\centering		
\scalebox{0.5}{							
\begin{tabular}{||c|c|c|c|c|}									
\hline									
Sample Cluster \#	&	GO ID	&	corr p-value	&	Gene Count	&	GO description	\\
\hline									
1	&	22403	&	5.93E-20	&	25/58	&	cell cycle phase	\\
:Gene Cluster 44	&	22402	&	3.78E-18	&	26/58	&	cell cycle process	\\
	&	279	&	1.77E-16	&	21/58	&	M phase	\\
	&	7049	&	8.78E-15	&	26/58	&	cell cycle	\\
	&	51301	&	9.84E-11	&	16/58	&	cell division	\\
	&	51726	&	1.12E-10	&	18/58	&	regulation of cell cycle	\\
	&	278	&	1.19E-10	&	17/58	&	mitotic cell cycle	\\
	&	6996	&	5.99E-10	&	27/58	&	organelle organization	\\
	&	16043	&	4.30E-08	&	33/58	&	cellular component organization	\\
	&	280	&	1.64E-07	&	12/58	&	nuclear division	\\
	&	7067	&	1.64E-07	&	12/58	&	mitosis	\\
	&	87	&	2.31E-07	&	12/58	&	M phase of mitotic cell cycle	\\
	&	48285	&	2.54E-07	&	12/58	&	organelle fission	\\
	&	6259	&	1.88E-06	&	15/58	&	DNA metabolic process	\\
	&	6974	&	6.49E-06	&	13/58	&	response to DNA damage stimulus	\\
	&	51321	&	1.11E-05	&	8/58	&	meiotic cell cycle	\\
	&	75	&	1.50E-05	&	8/58	&	cell cycle checkpoint	\\
	&	6281	&	3.75E-05	&	11/58	&	DNA repair	\\
	&	44260	&	4.41E-05	&	34/58	&	cellular macromolecule metabolic process	\\
	&	48522	&	9.05E-05	&	25/58	&	positive regulation of cellular process	\\
	&	65009	&	1.48E-04	&	18/58	&	regulation of molecular function	\\
	&	33554	&	1.69E-04	&	14/58	&	cellular response to stress	\\
	&	51276	&	1.87E-04	&	13/58	&	chromosome organization	\\
	&	79	&	2.05E-04	&	6/58	&	regulation of cyclin-dependent protein kinase activity	\\
	&	7126	&	2.42E-04	&	7/58	&	meiosis	\\
	&	51327	&	2.42E-04	&	7/58	&	M phase of meiotic cell cycle	\\
	&	51716	&	2.92E-04	&	17/58	&	cellular response to stimulus	\\
	&	50790	&	5.38E-04	&	16/58	&	regulation of catalytic activity	\\
	&	48518	&	6.09E-04	&	25/58	&	positive regulation of biological process	\\
	&	90304	&	7.63E-04	&	20/58	&	nucleic acid metabolic process	\\
	&	6139	&	1.03E-03	&	22/58	&	nucleobase, nucleoside, nucleotide and nucleic acid metabolic process	 \\
	&	9987	&	1.13E-03	&	54/58	&	cellular process	\\
	&	43170	&	1.50E-03	&	34/58	&	macromolecule metabolic process	\\
	&	51340	&	1.54E-03	&	6/58	&	regulation of ligase activity	\\
	&	7051	&	2.17E-03	&	5/58	&	spindle organization	\\
	&	44237	&	2.65E-03	&	38/58	&	cellular metabolic process	\\
	&	65003	&	3.28E-03	&	13/58	&	macromolecular complex assembly	\\
	&	34641	&	3.42E-03	&	23/58	&	cellular nitrogen compound metabolic process	\\
	&	51329	&	4.83E-03	&	6/58	&	interphase of mitotic cell cycle	\\
	&	6310	&	5.41E-03	&	6/58	&	DNA recombination	\\
	&	51325	&	6.05E-03	&	6/58	&	interphase	\\
	&	43933	&	6.96E-03	&	13/58	&	macromolecular complex subunit organization	\\
	&	6266	&	7.42E-03	&	3/58	&	DNA ligation	\\
	&	42127	&	7.43E-03	&	14/58	&	regulation of cell proliferation	\\
	&	48519	&	8.31E-03	&	22/58	&	negative regulation of biological process	\\
	&	6807	&	8.92E-03	&	23/58	&	nitrogen compound metabolic process	\\
\hline									
4	&	50851	&	4.41E-02	&	3/33	&	antigen receptor-mediated signaling pathway	\\
: Gene Cluster 109    &		&		&		&		\\
	&		&		&		&		\\
\hline									
5	&	44260	&	3.16E-14	&	107/206	&	cellular macromolecule metabolic process	\\
: Gene Cluster 1	&	43170	&	2.74E-11	&	110/206	&	macromolecule metabolic process	\\
	&	44237	&	4.72E-10	&	123/206	&	cellular metabolic process	\\
	&	43687	&	4.40E-09	&	52/206	&	post-translational protein modification	\\
	&	44238	&	1.76E-08	&	124/206	&	primary metabolic process	\\
	&	43412	&	1.11E-07	&	57/206	&	macromolecule modification	\\
	&	6464	&	1.47E-07	&	55/206	&	protein modification process	\\
	&	44267	&	3.12E-06	&	65/206	&	cellular protein metabolic process	\\
	&	8152	&	4.17E-06	&	128/206	&	metabolic process	\\
	&	50794	&	8.38E-06	&	131/206	&	regulation of cellular process	\\
	&	6468	&	1.05E-05	&	31/206	&	protein amino acid phosphorylation	\\
	&	90304	&	2.33E-05	&	49/206	&	nucleic acid metabolic process	\\
	&	10468	&	2.74E-05	&	77/206	&	regulation of gene expression	\\
	&	6796	&	4.94E-05	&	37/206	&	phosphate metabolic process	\\
	&	6793	&	4.94E-05	&	37/206	&	phosphorus metabolic process	\\
	&	16310	&	5.55E-05	&	33/206	&	phosphorylation	\\
	&	34641	&	8.94E-05	&	60/206	&	cellular nitrogen compound metabolic process	\\
	&	6139	&	1.23E-04	&	54/206	&	nucleobase, nucleoside, nucleotide and nucleic acid metabolic process	 \\
	&	31323	&	1.34E-04	&	89/206	&	regulation of cellular metabolic process	\\
	&	51171	&	1.47E-04	&	77/206	&	regulation of nitrogen compound metabolic process	\\
	&	10556	&	1.64E-04	&	74/206	&	regulation of macromolecule biosynthetic process	\\
	&	16071	&	1.84E-04	&	21/206	&	mRNA metabolic process	\\
	&	19219	&	2.31E-04	&	76/206	&	regulation of nucleobase, nucleoside, nucleotide and nucleic acid metabolic process	\\
	&	45449	&	2.36E-04	&	69/206	&	regulation of transcription	\\
	&	6807	&	2.73E-04	&	61/206	&	nitrogen compound metabolic process	\\
	&	50789	&	3.65E-04	&	131/206	&	regulation of biological process	\\
	&	60255	&	6.25E-04	&	81/206	&	regulation of macromolecule metabolic process	\\
	&	7165	&	6.60E-04	&	54/206	&	signal transduction	\\
	&	19538	&	1.09E-03	&	67/206	&	protein metabolic process	\\
	&	31326	&	1.24E-03	&	74/206	&	regulation of cellular biosynthetic process	\\
	&	80090	&	1.32E-03	&	83/206	&	regulation of primary metabolic process	\\
	&	19222	&	1.35E-03	&	89/206	&	regulation of metabolic process	\\
	&	9889	&	1.71E-03	&	74/206	&	regulation of biosynthetic process	\\
	&	16070	&	2.21E-03	&	34/206	&	RNA metabolic process	\\
	&	6357	&	6.65E-03	&	28/206	&	regulation of transcription from RNA polymerase II promoter	\\
	&	7049	&	7.26E-03	&	29/206	&	cell cycle	\\
\hline									
7	&	48102	&	7.16E-03	&	2/30	&	autophagic cell death	\\
:Gene Cluster 4    &		&		&		&		\\
	&		&		&		&		\\
\hline									
9	&	6955	&	5.56E-09	&	21/75	&	immune response	\\
: Gene Cluster 125	&	2376	&	7.82E-09	&	25/75	&	immune system process	\\
	&	9611	&	2.20E-05	&	16/75	&	response to wounding	\\
	&	6952	&	2.22E-05	&	17/75	&	defense response	\\
	&	6950	&	4.28E-05	&	28/75	&	response to stress	\\
	&	23052	&	5.59E-05	&	38/75	&	signaling	\\
	&	50896	&	8.44E-05	&	41/75	&	response to stimulus	\\
	&	6954	&	1.15E-04	&	12/75	&	inflammatory response	\\
	&	6935	&	3.37E-04	&	9/75	&	chemotaxis	\\
	&	42330	&	3.37E-04	&	9/75	&	taxis	\\
	&	40011	&	5.96E-04	&	13/75	&	locomotion	\\
	&	23033	&	1.55E-03	&	28/75	&	signaling pathway	\\
	&	9607	&	4.73E-03	&	12/75	&	response to biotic stimulus	\\
	&	22603	&	5.24E-03	&	10/75	&	regulation of anatomical structure morphogenesis	\\
	&	7165	&	7.87E-03	&	25/75	&	signal transduction	\\
	&	7166	&	9.01E-03	&	20/75	&	cell surface receptor linked signaling pathway	\\
\hline									
11	&		&		&		&		\\
(CLL Cluster)	&		&		&		&		\\
: Gene Cluster 102	&		&		&		&		\\
\hline									
\end{tabular}}								
\caption{\label{AllSampleClusterGO} Over-represented GO terms for each of the significant gene-clusters of sample-clusters 1, 5, 7, 9 (associated to DLBCL), 4 (associated to FL) and 11 (associated to CLL).}									 \end{table}		

\begin{table}			
\centering			
\scalebox{0.7}{			
\begin{tabular}{|l|l|}	
\hline\hline			
Clone name	 \\
\hline\hline			
*LyGDI=Rho GDP-dissociation inhibitor 2=RHO GDI 2; Clone=23	\\	*LyGDI=Rho GDP-dissociation inhibitor 2=RHO GDI 2; Clone=1240974	\\
*FLI-1=ERGB=ets family transcription factor; Clone=280882	\\	*FLI-1=ERGB=ets family transcription factor; Clone=1354062	\\
(Arp2/3 protein complex subunit p34-Arc (ARC34); Clone=1334980)	\\	(Unknown  UG Hs.28242  ESTs; Clone=1303641)	\\
(Aconitase=mitochondrial protein; Clone=1353272)	\\	(B-actin, 421-689; Clone=136)	\\
(B-actin,177-439; Clone=137)	\\	(Retinoblastoma-like 1 (p107); Clone=249725)	\\
(B-actin, 657-993; Clone=145)	\\	*actin=cytoskeletal gamma-actin; Clone=1240822	\\
*Similar to nuclear protein NIP45=potentiates NFAT-driven interleukin-4 transcription; Clone=512953	 \\*actin=cytoskeletal gamma-actin; Clone=588637	\\
*Adenine nucleotide translocator 2; Clone=291660	\\	*Adenine nucleotide translocator 2; Clone=1241102	\\
*Calmodulin 1 (phosphorylase kinase, delta); Clone=549080	\\	
\hline\hline			
\end{tabular}}			
\caption{\label{GeneCluster11List}List of clones in gene-cluster 102, which corresponds to the most significant gene-cluster for sample-cluster 11 associated to CLL.} 			
\end{table}

\end{appendix}


\end{document}